\documentclass[]{interact}

\usepackage{epstopdf}
\usepackage[caption=false]{subfig}

\usepackage{amsfonts, amssymb, amsmath, bm}
\usepackage{graphicx}
\usepackage{stackrel}
\usepackage{csquotes}
\usepackage{ulem}
\usepackage{mathptmx}
\usepackage{multirow, array, tabularx}
\usepackage{siunitx}
\usepackage{pgfplots}
\usepackage[mathscr]{euscript}

\usepackage[numbers,sort&compress]{natbib}
\bibpunct[, ]{[}{]}{,}{n}{,}{,}
\renewcommand\bibfont{\fontsize{10}{12}\selectfont}

\usepackage[colorlinks=true, allcolors=blue]{hyperref} 

\renewcommand{\vec}[1]{\bm{#1}}

\newcommand{\Pra}{{\ensuremath{\text{Pr}}}}

\newcommand{\vI}{\ensuremath{\vec{v}_\text{I}}}
\newcommand{\xb}{\ensuremath{x_\text{b}}}
\newcommand{\xI}{\ensuremath{x_\text{I}}}

\newcommand{\Rey}{\ensuremath{\text{Re}}}

\newcommand{\Fdrag}{\ensuremath{\vec F_{\text{drag}}}}
\newcommand{\Fb}{\ensuremath{\vec F_{\text{b}}}}
\newcommand{\cs}{c_\text{s}}
\newcommand{\Fba}{\ensuremath{\vec F_{\text{b},\alpha}}}
\newcommand{\Pna}{\ensuremath{P_{0,\alpha}}}

\newcommand{\Pha}{\ensuremath{P_{\text{h},\alpha}}}
\newcommand{\HTfd}{\ensuremath{H^\text{f.d.}_T}}
\newcommand{\Hufd}{\ensuremath{H^\text{f.d.}_u}}
\newcommand{\rhoa}{\ensuremath{\rho_{\alpha}}}
\newcommand{\fa}{\ensuremath{f_{\alpha}}}
\newcommand{\ua}{\ensuremath{\vec u_{\alpha}}}
\newcommand{\ub}{\ensuremath{\vec u_{\beta}}}
\newcommand{\uc}{\ensuremath{u_\text{c}}}
\newcommand{\uinlet}{\ensuremath{u_\text{inlet}}}
\newcommand{\umax}{\ensuremath{u_{\text{max}}}}
\newcommand{\phia}{\ensuremath{\varphi_{\alpha}}}
\newcommand{\phib}{\ensuremath{\varphi_{\beta}}}
\newcommand{\chia}{\ensuremath{\chi_{\alpha}}}
\newcommand{\chib}{\ensuremath{\chi_{\beta}}}
\newcommand{\taua}{\ensuremath{\vec \tau_{\alpha}}}
\newcommand{\lambdaa}{\ensuremath{\lambda_{\alpha}}}
\newcommand{\Austar}{\ensuremath{{A_u}^\ast}}
\newcommand{\ATstar}{\ensuremath{{A_T}^\ast}}
\newcommand{\Ta}{\ensuremath{T_{\alpha}}}
\newcommand{\Tb}{\ensuremath{T_{\beta}}}

\newcommand{\Twall}{\ensuremath{T_\text{wall}}}
\newcommand{\Tcylinder}{\ensuremath{T_\text{cylinder}}}
\newcommand{\Tinlet}{\ensuremath{T_\text{inlet}}}
\newcommand{\Tadiab}{\ensuremath{T_\text{ad}}}
\newcommand{\Tm}{\ensuremath{T_\text{m}}}
\newcommand{\Mw}{\ensuremath{\bar{M}_\text{w}}}
\newcommand{\ha}{\ensuremath{h_{\alpha}}}

\newcommand{\qa}{\ensuremath{\vec q_{\alpha}}}
\newcommand{\qaI}{\ensuremath{\vec q_{\alpha,\text{I}}}}
\newcommand{\Ph}{\ensuremath{P_{\text{h}}}}
\newcommand{\PhaI}{\ensuremath{P_{\text{h},\alpha,\text{I}}}}
\newcommand{\SIa}{\ensuremath{S_{\alpha,\text{I}}}}

\newcommand{\Yka}{\ensuremath{Y_{k,\alpha}}}

\newcommand{\jka}{\ensuremath{\vec j_{k,\alpha}}}
\newcommand{\jk}{\ensuremath{\vec j_{k}}}
\newcommand{\jkaI}{\ensuremath{\vec j_{k,\alpha,\text{I}}}}

\newcommand{\tauaI}{\ensuremath{\vec \tau_{\alpha,\text{I}}}}

\newcommand{\blue}[1]{\textcolor{black}{#1}}

\newcommand{\hideimage}[1]{#1}
\newcommand{\hide}[1]{}

\begin{document}


\title{A low-Mach phase field-lattice Boltzmann-finite difference model for reactive gas flows propagating through complex shaped particle assemblies}

\author{
\name{Reza Namdar\textsuperscript{a}, Mohammad Norouzi\textsuperscript{a}, Fathollah Varnik\textsuperscript{a*}\thanks{*Corresponding author. Email: fathollah.varnik@rub.de} }
\affil{\textsuperscript{a}Interdisciplinary Centre for Advanced Materials Simulation (ICAMS), Ruhr-University Bochum, Universitätsstraße 150, 44801 Bochum, Germany}
}

\maketitle

\begin{abstract}
The present study provides a systematic derivation of a phase-field version of the momentum, mass and heat transport equations, while accounting for chemical reactions in the fluid phase. To achieve this goal, the volume averaging technique is used to reformulate the conservation equations in the presence of multiple phases and their respective diffuse interfaces. It is shown that the structure of the multiphase/diffuse interface version of the conservation equations is very similar to the original single phase/sharp interface formulation. The multiphase character of the problem is accounted for the coupling terms, which act at the interface between adjacent phases. For the special case of a reactive fluid in contact with an inert solid, two coupling parameters are introduced, which control the exchange of momentum and heat at the interface. For numerical solver, a low-Mach number hybrid lattice Boltzmann-finite difference-phase field (LB-FD-PF) framework is developed and implemented in the open source software OpenPhase Academic. Chemical reactions of reactive flows are included into the model by coupling OpenPhase Academic with the open-source chemical kinetics software CANTERA, which delivers details of the chemical reaction mechanisms and the necessary thermodynamic and transport properties of the reacting chemical species. The model is thoroughly validated against alternative numerical simulations of reactive flows as well as experiments.
\end{abstract}


\section{Introduction}

Exchange of heat and momentum with the solid phase is an issue of central importance in the flow of reactive fluids through an assembly of solid particles (packed beds)~\cite{Anderson1984}. Since this process occurs across the fluid-solid interface, an adequate numerical modelling of heat and momentum transfer through interfaces is of utmost importance in this area of research. The fluid-solid interface has a thickness of a few atomic diameters, where properties such as density and heat capacity change in a continuous manner from those of a solid to their corresponding values in the fluid phase. Due to this extremely small linear dimension, the thickness of the interface is assumed to be zero in continuum fluid mechanics applications. As a consequence of this {\it sharp interface} (SI) approach, physical properties undergo a sudden change at the interface. In simple cases, where exact analytic solutions can be worked out, discontinuous changes at the interface are accounted for via appropriate boundary conditions. In more complex situations, however, one is often dependent on numerical methods, where space and time are discretized and the underlying partial differential equations are solved using sufficiently small integration steps. In almost all these approaches, imposing boundary conditions --- while keeping a high accuracy --- requires computation of gradients and almost inevitably leads to a numerical interface, the extension of which depends on the details of the underlying approximation scheme~\cite{Poinsot2005}. A common challenge here is dealing with sharp edges and corners of objects, where spatial derivatives become large and also strongly depend on the direction along which derivative is evaluated. To avoid possible numerical instabilities, this often requires tedious numerical treatments to account for the specific geometry at corners/edges. Moreover, in cases where interfaces are not static but change with time, methods which aim at implementing a sharp fluid-solid boundary often need to track the time evolution of the complex interface~\cite{Shaikh2018}.

As an interesting alternative, diffuse interface (DI) methods --- where properties change smoothly across the fluid-solid interface --- provide the possibility to circumvent the need for interface tracking, while at the same time allowing a softening of sharp edges~\cite{Anderson1998}. Some examples for DI-based numerical approaches are volume of fluid method, level set method and the phase field method~\cite{Worner2012,Liu2022,Liu2025}. Taking advantage of our experience in lattice Boltzmann (LB) modelling of fluid dynamical phenomena~\cite{Varnik2007,Varnik2008,Ayodele2011,Ayodele2015} and the phase field (PF) approach~\cite{Schiedung2017,Schiedung2018,Schiedung2020,Vakili2020}, we recently developed a monolithic simulation approach, which combines these methods to study heat and mass transfer in fluid flows through particle beds~\cite{Namdar2023}. Here, we address the mathematical foundation for that work and present a systematic derivation of diffuse interface-based equations for mass, momentum and energy in the presence of chemical reactions occurring in the fluid phase.

A key step along this line is the choice of an adequate coupling scheme between the fluid and solid phases to account for the variation of physical properties and their exchange across the interface. Even though reasonable choices can be made by physically motivated empirical assumptions, a systematic approach proves useful, as it can provide answers also in intuitively less clear situations. In this context, the so-called volume averaging technique provides a transparent way to derive partial differential equations for a domain containing different phases~\cite{Drew1983}. One of the early works on the volume averaging method goes back to D.A. Drew, who proposed this technique to derive momentum and mass balance equations for two phase flows~\cite{Drew1983}.

Recently, we have applied this technique to investigate coupling between the phase field method and the Navier-Stokes equation for an isothermal flow~\cite{Beckermann1999,Subhedar2015}. Here we extend the methodology to account for heat transfer as well as chemical reactions in the fluid phase with a focus on momentum and energy exchange at a diffuse fluid-solid interface. This extension modifies the energy equation at a diffuse fluid-solid interface. The primary motivation for adopting the phase field approach lies in its capabilities to deal with complex geometries and investigating the kinetics of phase transformation in multiphase systems for future research. For the sake of completeness, we first give below a brief survey of the volume averaging technique. Some of the basic equations introduced here will simplify the derivations of transport equation in section~\ref{subsec:DI-formulation}.

\section{Volume averaging technique}
\label{sec:volume-average}

\begin{figure}
\centering
\hideimage{\includegraphics[width=\linewidth]{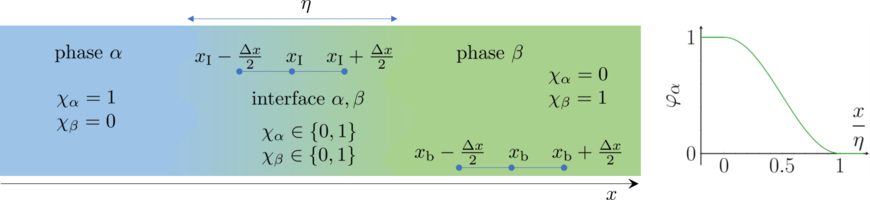}\hspace*{5mm}}
	\caption{Left: A schematic two dimensional view of the two phases $\alpha$ and $\beta$ in contact with each other. The physical phase is assumed to be uniform along the vertical direction but varies along the horizontal one, $x$. If $\partial\chia/\partial x \neq 0$ at a given point in space, it directly follows that the considered point belongs to the interface domain. The point $\xI$ illustrates such a case. In the bulk region (see, e.g., the point $\xb$), on the other hand, a point is surrounded by the same physical phase so that the gradient of the characteristic function ($\chia$ or $\chib$) vanishes at that point. Thus, $\nabla\chi$ can be viewed as an indicator of the interface domain. Right: Variation of the phase-field function, $\phia=\langle \chia \rangle$, across the interface domain. The smoothness of $\phia$ implies that the linear size of the volume averaged domain must be smaller than the interface thickness, $\Delta x\ll\eta$. For a better visibility, however, $\Delta x$ is chosen to be roughly equal to $\frac{1}{2}\eta$ in this plot.}
	\label{fig:phase-field}
\end{figure}

The central quantity in the volume averaging technique is the characteristic function $\chia$, which gives information about the presence or absence of the phase $\alpha$ on the subgrid scale,
\begin{equation}
\chia(\vec r)=
    \begin{cases}
    1&\;\;\;\text{if phase $\alpha$ is present at }\vec r\;\;\\	
    0&\;\;\;\text{else}
    \end{cases}
\end{equation}
It is assumed that any point $\vec r$ in the subgrid scale contains exactly one physical phase, whereby void space is viewed as a \enquote{phase} either. Thus, if $\chia(\vec r)=1$, then $\chi_\beta(\vec r)=0$ for all other phases $\beta\neq \alpha$. It then follows that
\begin{equation}
\sum_{\alpha}\chia(\vec r)=1
\label{eq:sum-over-chi-alpha-is-one}
\end{equation}
for an arbitrary point $\vec r$ in space. The characteristic function $\chi$ provides information about the spatial distribution of phases on a scale much smaller than the numerical grid resolution (i.e., on a sub-grid scale). Let now $\Delta x$, $\Delta y$ and $\Delta z$ be the actual grid scales along $x$, $y$ and $z$ coordinate-directions. For a smooth and differentiable but otherwise arbitrary function $f$, the volume average inside a volume element of size $\Delta V=\Delta x\Delta y\Delta z$ centered at the point $\vec r=(x,y,z)$ is defined via 
\begin{equation}
\langle f \rangle (\vec r)=\frac{1}{\Delta V} \int_{x-\Delta x/2}^{x+\Delta x/2} \,dx'  \int_{y-\Delta y/2}^{y+\Delta y/2} \,dy'  
\int_{z-\Delta z/2}^{z+\Delta z/2}\,dz' f (\vec r').
\label{eq:def-volume-average}
\end{equation}
In the next step, one defines the so-called phase-field function, $\phia$, as the fraction of $\Delta V$ which is occupied by the phase $\alpha$,
\begin{equation}
\phia(\vec r)=\langle \chia \rangle (\vec r)  \quad    \text{(subgrid scale)}.
\label{eq:phase-field-definition}
\end{equation}
As a result of this averaging process, noticeable changes of $\phia$ can occur only if $\vec r$ changes by distances comparable to or larger than $\Delta x$, $\Delta y$, or $\Delta z$, see Fig.\ref{fig:phase-field}. Applying the averaging operator to Eq.~(\ref{eq:sum-over-chi-alpha-is-one}) yields, $1=\langle 1\rangle=\langle \sum_\alpha  \chia \rangle= \sum_\alpha \langle \chia \rangle$. We now use Eq.~(\ref{eq:phase-field-definition}), and arrive at an important constraint for the phase field function,
\begin{equation}
\sum_\alpha \phia (\vec r)=1   \quad    \text{(grid scale)}.
\label{eq:sum-varphialpha}
\end{equation}
Next we introduce $\fa$ as the value of a property $f$ inside the phase $\alpha$. Some typical examples for $f$ in the present problem are density, $f\equiv \rho$, temperature $f\equiv T$, flow velocity, $\vec f\equiv \vec u$ or mass current density $\vec f\equiv \rho\vec u$. For any point in space, one can write $\chia(\vec r)\,f(\vec r)=\chia(\vec r)\,\fa(\vec r)$. To see this, we recall that $\chia(\vec r)$ can take only one of the two discrete values of 0 or 1. Obviously, the equality holds for $\chia=0$ for any $f$. In the other case that $\chia=1$, the point $\vec r$ is occupied by the physical phase $\alpha$, and thus $f(\vec r)=\fa(\vec r)$. Using this fact, we evaluate the volume average of the product $\chia\,f$ as follows,
\begin{align}
\langle f\, \chia \rangle (\vec r)&=\frac{1}{\Delta V} \int_{x-\Delta x/2}^{x+\Delta x/2} \,dx'  \int_{y-\Delta y/2}^{y+\Delta y/2} \,dy'  
\int_{z-\Delta z/2}^{z+\Delta z/2}\,dz'\,  \fa (\vec r')\, \chia (\vec r'),\label{eq:volume-average-fchia-1}\\
& = \fa(\vec r)\, \frac{1}{\Delta V} \int_{x-\Delta x/2}^{x+\Delta x/2} \,dx'  \int_{y-\Delta y/2}^{y+\Delta y/2} \,dy'  
\int_{z-\Delta z/2}^{z+\Delta z/2}\,dz'\, \chia (\vec r'),\label{eq:volume-average-fchia-2}\\
&=\fa(\vec r)\, \langle \chia \rangle (\vec r).\label{eq:volume-average-fchia-3}
\end{align}
In going from Eq.~(\ref{eq:volume-average-fchia-1}) to Eq.~(\ref{eq:volume-average-fchia-2}), we used the fact that $\fa$ is resolved only on the grid scale and can thus be assumed constant inside the volume-averaging domain $\Delta V$. 
Using Eq.~(\ref{eq:phase-field-definition}), we rewrite Eq.~(\ref{eq:volume-average-fchia-3}) as,
\begin{equation}
	\langle  \chia\, f\rangle (\vec r)=\phia (\vec r) \fa (\vec r).
	\label{eq:def-volume-average-fchia}
\end{equation}
In the diffuse interface model to be worked out in this work, all physical properties are resolved on the grid scale. The information about the subgrid scale influences only the value of $\phia$. The average value of a property $f$ at point $\vec r$ (i.e. inside a volume $\Delta V$ centered at $\vec r$) is a weighted sum over the values, which $f$ takes within individual physical phases present at $\vec r$. To see this, we start with Eq.~(\ref{eq:def-volume-average-fchia}) and perform the sum over all phases. This yields,
\begin{align}
\sum_\alpha \phia (\vec r) \fa (\vec r) &= \sum_\alpha \langle \chia\, f\rangle (\vec r),\label{eq:sum-falpha-varphialpha-1}\\
&=\langle \Big( \sum_\alpha \chia\Big) f \rangle (\vec r),\label{eq:sum-falpha-varphialpha-2}\\
&=\langle f \rangle (\vec r),
\label{eq:sum-falpha-varphialpha}
\end{align}
where we used the fact that the order of the summation over phases and the subgrid scale integration over the volume $\Delta V$ can be changed (Eq.~(\ref{eq:sum-falpha-varphialpha-1}) $\to$ Eq.~(\ref{eq:sum-falpha-varphialpha-2})). We then used Eq.~(\ref{eq:sum-over-chi-alpha-is-one}), which expresses the fact that, in contrast to $\Delta V$, inside which different phases may coexist, a sub-grid point contains exactly one physical phase. 

It is worth elucidating the meaning of Eq.~(\ref{eq:sum-falpha-varphialpha}) in a specific example. For this purpose, we consider $f\equiv \rho$ in a system containing two phases only, a solid phase and a fluid phase. Let $\alpha=\text{fluid}$, then $\langle \rho \rangle$ gives the value of density obtained by weighted averaging over the solid and fluid densities inside the volume element $\Delta V$ around the point $\vec r$. Obviously, if $\phi_\text{fluid}=1$ the entire volume element at $\vec r$ is filled with fluid so that $\langle \rho \rangle =\rho_\text{fluid}$. In the opposite case of $\phi_\text{fluid}=0$, on the other hand, $\langle \rho\rangle =\rho_\text{solid}$. For intermediate values, $0<\phi_\text{fluid}<1$, one obtains for $\langle \rho \rangle$ a linear combination of $\rho_\text{fluid}$ and $\rho_\text{solid}$.

Following a similar reasoning as in the above lines, it is straightforward to derive the vectorial version of Eq.~(\ref{eq:def-volume-average-fchia}), which reads,
\begin{equation}
	\langle  \chia\, \vec f \rangle (\vec r)=\phia (\vec r) \vec \fa (\vec r).
	\label{eq:def-volume-average-chia-fvec}
\end{equation}
If one replaces the property $f$ on the lhs of Eq.~(\ref{eq:volume-average-fchia-1}) by the product of two properties, say $fg$, performing all the above steps yields in this case,
\begin{equation}
	\langle  \chia\, f\, g\rangle (\vec r)=\phia (\vec r) \fa (\vec r)\, g_\alpha(\vec r),
	\label{eq:def-volume-average-fgchia}
\end{equation}
where we again used the fact that $f$ and $g$ are not resolved on the subgrid scale.

For simplicity of notation, we will omit in the remaining of this work the dependence on the variables $\vec r$ (and $t$, see time derivatives below) but will keep this dependence in mind. To proceed further, we introduce a differential operator, $\partial_x\equiv \frac{\partial}{\partial x}$, which acts on the grid scale, apply it to Eq.~(\ref{eq:def-volume-average}) and obtain~\cite{Subhedar2015}
\begin{align}
\langle \partial_x f \rangle   & = \partial_x \langle f \rangle,
\label{eq:average-of-derivative-wrt-x}
\end{align}
where we used the fact that the order of differentiation and volume averaging can be exchanged since they act on different scales (see, e.g.,~\cite{Drew1983,Subhedar2015}). Relations similar to Eq.~(\ref{eq:average-of-derivative-wrt-x}) are readily obtained for derivative with respect to $y$ and $z$ coordinates. Using vectorial notation, one can thus write,
\begin{align}
\langle \nabla f \rangle  & = \nabla \langle f \rangle.
\label{eq:average-of-nabla-f}
\end{align}
Moreover, since the order of time derivative ($\partial_t \equiv \frac{\partial }{\partial t}$) and spatial integration can be exchanged, one has,
\begin{align}
\langle \partial_t f \rangle  & = \partial_t \langle f \rangle.
\label{eq:average-of-derivative-wrt-t}
\end{align}
As an important relation to be used in the next section, we also address the volume average of $\chia\,\nabla f$. 
Using the product rule of differentiation and Eq.~(\ref{eq:average-of-derivative-wrt-x}), we can write,
\begin{equation}
\langle \chia \nabla f \rangle =\nabla \langle \chia f \rangle - \langle f \nabla \chia \rangle
\label{eq:chia-nablaf-1}
\end{equation}
Inserting Eq.~(\ref{eq:def-volume-average-fchia}) into Eq.~(\ref{eq:chia-nablaf-1}) then yields
\begin{equation}
\langle \chia \nabla f \rangle =\nabla (\phia \fa) - \langle f \nabla \chia \rangle.
\label{eq:average-of-chi-nablaf}
\end{equation}
It is easy to repeat the above procedure for the case of a vector function $\vec f$ and show that 
\begin{align}
\langle \chia\, \nabla \cdot \vec f \rangle  & = \nabla \cdot (\phia\, \vec \fa) - \langle \vec f \cdot \nabla \chia \rangle.
\label{eq:average-of-chi-nablafvec}
\end{align}
Similarly, in the case of a second rank tensor $\mathbb{T}$, one finds
\begin{align}
\langle \chia\, \nabla \cdot \mathbb{T} \rangle  & = \nabla \cdot (\phia\, \mathbb{T}_\alpha) - \langle  \mathbb{T} \cdot \nabla \chia \rangle.
\label{eq:average-of-chi-nabla-second-rank-tensor}
\end{align}

As to the time derivative of a function, multiplication with $\chia$ and volume averaging leads to $\langle \chia \partial_t f \rangle$, which can then be rewritten as
\begin{align}
\langle \chia \partial_t f \rangle & = \langle  \partial_t (\chia f) \rangle - \langle f\, \partial_t \chia\rangle, \label{eq:average-of-chi-df-dt-1}\\
&=\partial_t \langle \chia\, f \rangle + \langle f\, \vI\cdot \nabla \chia \rangle, \label{eq:average-of-chi-df-dt-0}
\end{align}
where $\vI$ is the interface velocity. In going from Eq.~(\ref{eq:average-of-chi-df-dt-1}) to Eq.~(\ref{eq:average-of-chi-df-dt-0}), we used $\frac{d\chia}{dt}=\partial_t \chia + \vI\cdot\nabla\chia=0$, which reflects the fact that a detector moving with the interface does not observe any change of $\chia$, since its neighborhood remains the same. Using Eq.~(\ref{eq:def-volume-average-fchia}), one can rewrite Eq.~(\ref{eq:average-of-chi-df-dt-0}) as
\begin{equation}
\langle \chia \partial_t f \rangle = \partial_t (\phia\, \fa) + \langle f\, \vI\cdot \nabla \chia \rangle.
\label{eq:average-of-chi-df-dt}
\end{equation}

In the case of a vectorial function, $\vec f$, Eq.~(\ref{eq:average-of-chi-df-dt}) holds for each component of the function separately. Hence, one can write,
\begin{equation}
\langle \chia \partial_t \vec f \rangle = \partial_t (\phia\, \vec \fa) + \langle \vec f\, \vI\cdot \nabla \chia \rangle.
\label{eq:average-of-chi-dfvec-dt}
\end{equation}

Results obtained in this section will prove useful below in deriving volume averaged equations for momentum balance, energy and species transport in the presence of chemical reactions.

\section{Model}
In this section, we first describe the sharp-interface version of partial differential equations for a reactive fluid flow and then work out the diffuse-interface version of the same equations by applying the volume averaging technique introduced in section \ref{sec:volume-average}.

\subsection{Sharp-interface equations for reactive flows}
In combustion problems through packed beds, Mach number is relatively low and density variations occur due to temperature changes. For this reason, we employ a thermal compressible model known as Low Mach Number Approximation (LMNA) which accounts for temperature changes and ignores sound waves. Within this approximation, the conservation laws of mass, momentum and energy, as well as change of species concentration are expressed by the following partial differential equations,
\begin{align}
\partial_t \rho +  \nabla\cdot (\rho \bm{u}) &= 0,  \label{eq:mass-cons}\\
\partial_t (\rho \bm{u}) +  \nabla\cdot (\rho \vec u \vec u) &= -  \nabla P_\text{h}+  \nabla\cdot \bm{\tau} + \Fb, \label{eq:momentum-cons}\\
\partial_t (\rho Y_k)  +  \nabla\cdot(\rho \bm{u} Y_k) & = - \nabla\cdot \bm{j}_k  + \dot{\omega}_k, \label{eq:species-cons} \\
\partial_t  (\rho h) +  \nabla\cdot(\rho \bm{u} h) &= - \nabla \cdot \bm{q}  + \partial_t P_0 + \dot{Q}, \label{eq:energy-cons}\\
\rho(\vec r, t) &= \frac{M_\text{w}\, P_0(t)}{R\,T(\vec r, t)} \label{eq:rho-in-terms-of-P0-and-T}.
\end{align} 
In the above equations, $\rho$, $\bm{u}$, $\bm{\tau}$ and $\Fb$ are fluid density, fluid velocity, viscous stress tensor and external body force, respectively. In Eq.~(\ref{eq:momentum-cons}), $\vec u \vec u$ is a dyadic product, i. e., a symmetric second rank tensor whose components are $u_i u_j$, with $i,j\in\{x,y,z\}$. The first term on the rhs of this equation is the gradient of hydrodynamic pressure fluctuations, $P_\text{h}=P(\vec r, t)-P_0(t)$, where $P(\vec r, t)$ is the actual pressure and $P_0(t)$ represents a pressure, which would establish if the fluid would be at rest. $P_0$ is homogeneous in space but can vary with time due to  variations of temperature, $T$. 
It is introduced to account for the fact that, in the low Mach number limit considered in this work, sound propagation damps pressure gradients to a large extent. Thus, at each time instant $t$, spatial variations of $P(\vec r, t)$ are rather small and can be regarded as  spatial fluctuations of a uniform field, $P_0(t)$. Hence, $P_h(\vec r, t)$ represents these small spatial pressure variations. To calculate $P_0(t)$ in terms of the quantities available in the simulation, we consider the fluid to be an ideal gas mixture and write for the total mass inside the simulation domain $P_0=\frac{R}{\Mw}\rho(\vec r, t) T(\vec r, t)$, where $R=8.314$ J/(mol K) is the  universal gas constant and $\Mw$ is the mean molar mass of the mixture. Since both density and temperature vary in space in the present problem, the question arises as how to calculate a homogeneous pressure from these fields. To solve this problem, we use the conservation of mass inside the system as a guideline, and write for the total mass inside the simulation domain, $m=\int \rho(\vec r, t)dV=\Mw P_0 \int \frac{dV}{T}$. In the last step, we used the fact that $P_0$ is spatially uniform and can thus be taken out of the integral over space. Rewriting this equation yields $P_0=\frac{mR}{\Mw}/\int \frac{dV}{T}$. $Y_k$, $\bm{j}_k$ and $\dot{\omega}_k$ in Eq.~(\ref{eq:species-cons}) represent the mass fraction of the $k$-th chemical species, diffusion flux and production/destruction of species $k$, respectively. In Eq.~(\ref{eq:energy-cons}), $h$, $\bm{q}$ and  $\dot{Q}$ are enthalpy per unit mass, heat flux and heat source term due to chemical reactions, respectively. Equation~(\ref{eq:rho-in-terms-of-P0-and-T}) couples the local density to the local temperature.

The viscous stress tensor, $\vec \tau$, the diffusion flux for the $k$-th chemical species, $\vec j_k$, and heat diffusion flux $\vec q$ are defined via:
\begin{align}
\vec \tau = \mu [ \nabla\vec u + ( \nabla\vec u)^T -& \frac{2}{d}( \nabla\cdot\vec u)\vec I]+\zeta ( \nabla\cdot\vec u)\vec I,  \label{eq:shear-stress}\\
\jk = -\rho \mathscr{D}_k &  \nabla Y_k, \label{eq:diffusion-flux}\\
\vec q = -\lambda  \nabla T &- \sum_{k=1}^{N} \jk h_k.  \label{eq:diffusion-heat-flux}
\end{align}
where $d$ is the dimension of space and $\mu$ and $\zeta$ are the dynamic shear and bulk viscosities, respectively. The superscript $T$ indicates the transpose operation. $\vec I$ is the identity tensor. $\mathscr{D}_k$,  $\lambda_k$ and $h_k$ denote mass diffusion coefficient, thermal conductivity and specific enthalpy of species $k$, respectively.

\subsection{Diffuse interface formulation} \label{subsec:DI-formulation}

Equations~(\ref{eq:shear-stress})-(\ref{eq:diffusion-heat-flux}) close the set of equations (\ref{eq:mass-cons})-(\ref{eq:rho-in-terms-of-P0-and-T}) and allow their solution, once boundary conditions are completely characterized. This introduces a new set of equations regarding the behavior of physical quantities at sharp interfaces. In this work, we choose a slightly different path and introduce a diffuse interface version of the above equations, which allows the change of properties across the fluid-solid boundary in a continuous manner.

To account for the presence of a diffuse fluid-solid boundary in equations~(\ref{eq:mass-cons})-(\ref{eq:energy-cons}), we multiply these equations with a characteristic function $\chi_{\alpha}$, which indicates the presence of a phase $\alpha$ (e.g., fluid phase), and perform the volume averaging procedure described in section~\ref{sec:volume-average}. 

In the next subsection (\ref{sec:DI-version-of-continuity-and-NS-equations}), this procedure is applied to the mass and momentum conservation equations~(\ref{eq:mass-cons}) and (\ref{eq:momentum-cons}). This essentially recapitulates the work published in~\cite{Subhedar2015} and is given here for the sake of completeness. The novel part of the work is then given in the subsequent sections, where transport equations for heat and individual chemical species are addressed formally (subsections~\ref{sec:Species-equations-at-a-diffuse-interface} and ~\ref{sec:Species-equations-at-a-diffuse-interface}) and then examined via computer simulations.

\subsubsection{DI-version of the continuity and NS-equations}
\label{sec:DI-version-of-continuity-and-NS-equations}
As a first step, we give here the derivation of volume averaged continuity equation. For this purpose, we multiply the mass conservation equation~(\ref{eq:mass-cons}) with the characteristic function $\chia$ and take the volume average. This yields
\begin{align}
\langle \chia\partial_t  \rho \rangle + \langle  \chia  \nabla\cdot (\rho \bm{u}) \rangle &= 0 .
\label{eq:vol-average-over-chia-x-mass-conservation-equation}
\end{align}
The first term on the lhs of Eq.~(\ref{eq:vol-average-over-chia-x-mass-conservation-equation}) can be rewritten via Eq.~(\ref{eq:average-of-chi-df-dt}), where we identify $f\equiv\rho$. For the second term, we use Eq.~(\ref{eq:average-of-chi-nablafvec}) and set $\vec f \equiv \rho\vec u$. After rearrangement of terms, one then arrives at
\begin{align}
\partial_t (\phia \rhoa)  +   \nabla \cdot(\phia\rhoa\ua) - \langle \rho(\vec u-\vI) \cdot\nabla \chia\rangle &=0.
\label{eq:volume-averaged-continuity-equation}
\end{align}
The volume-averaged version of the momentum balance equation Eq.~(\ref{eq:momentum-cons}) reads,
\begin{equation}
\langle  \chia  \partial_t  (\rho \vec u)  \rangle + \langle  \chia \nabla\cdot(\rho \vec u \vec u)  \rangle = -\langle \chia \nabla \Ph \rangle + \langle \chia \nabla\cdot \vec\tau\rangle  + \langle \chia \Fb \rangle.
\label{eq:volume-average-of-chi-x-NS-equation}
\end{equation}
We use Eq.~(\ref{eq:average-of-chi-dfvec-dt}) with $\vec f=\rho\vec u$ to rewrite the first term on the lhs of Eq.~(\ref{eq:volume-average-of-chi-x-NS-equation}). The tensorial expressions on the lhs and rhs of this equation are rewritten using Eq.~(\ref{eq:average-of-chi-nabla-second-rank-tensor}), where we set $\mathbb{T}=\rho\vec u\vec u$ (lhs) and $\mathbb{T}=\vec \tau$ (rhs), respectively. For the pressure term on the rhs we make use of Eq.~(\ref{eq:average-of-chi-nablaf}). We thus obtain
\begin{align}
&\partial_t  (\phia\rhoa \ua)+ \langle \rho \vec u\, \vI\cdot\nabla\chia\rangle + \nabla \cdot (\phia\rhoa\ua\, \ua) - \langle\rho\vec u\, \vec u\cdot\nabla\chia \rangle\nonumber\\
&=-\nabla(\phia\Pha) +\langle \Ph \nabla \chia \rangle + \nabla\cdot (\phia\taua) - \langle \vec \tau \cdot \nabla \chia \rangle + \phia \Fba,
\label{eq:DI-version-of-NS-equation-1}
\end{align}
where we used Eq.~(\ref{eq:def-volume-average-chia-fvec}) for the body force term on the rhs of Eq.~(\ref{eq:volume-average-of-chi-x-NS-equation}). The lhs of Eq.~(\ref{eq:DI-version-of-NS-equation-1}) can be considerably simplified by applying the product rule to the first term and using Eq.~(\ref{eq:volume-averaged-continuity-equation}). Rearranging terms then yields,
\begin{align}
\text{lhs of Eq.~(\ref{eq:DI-version-of-NS-equation-1})}&=\phia \rhoa \partial_t \ua + \ua [-\nabla \cdot (\phia\rhoa\ua) + \langle \rho (\vec u -\vI)\cdot\nabla \chia\rangle ]\nonumber\\
&+ \langle \rho \vec u (\vI-\vec u)\cdot\nabla\chia \rangle +\phia\rhoa\ua\cdot\nabla\ua + \ua\nabla\cdot(\phia\rhoa\ua) \nonumber\\
&=\phia\rhoa (\partial_t \ua + \ua\cdot\nabla\ua) - \langle \rho(\vec u-\ua)(\vec u-\vI)\cdot\nabla\chia \rangle\label{eq:volume-averaged-NS-equation-lhs-1}\\
&=\phia\rhoa (\partial_t \ua + \ua\cdot\nabla\ua) + {\cal{O}}(|\delta \vec u|)^2.\label{eq:volume-averaged-NS-equation-lhs-2}
\end{align}
In going from Eq.~(\ref{eq:volume-averaged-NS-equation-lhs-1}) to Eq.~(\ref{eq:volume-averaged-NS-equation-lhs-2}), we neglected terms of the second order in velocity differences at the interface. This is plausible since, due to the action of solid drag, fluid velocity in the proximity of a solid interface is expected to be close to the interface velocity so that velocity differences appearing on the rhs of Eq.~(\ref{eq:volume-averaged-NS-equation-lhs-1}) are small compared to the fluid velocity $\ua$. As to the rhs of Eq.~(\ref{eq:DI-version-of-NS-equation-1}), we apply the product rule and write
\begin{align}
\text{rhs of Eq.~(\ref{eq:DI-version-of-NS-equation-1})}&=\phia (-\nabla \Pha + \nabla \cdot \vec \taua +\Fba) - \Pha \nabla \phia + \langle \Ph \nabla \chia \rangle\nonumber\\
&+ \vec \taua \cdot \nabla \phia -\langle\vec \tau \cdot \nabla \chia \rangle.
\label{eq:DI-version-of-NS-equation-rhs-1}
\end{align}
Terms in the bracket on the rhs of Eq.~(\ref{eq:DI-version-of-NS-equation-rhs-1}) are the familiar pressure gradient, viscous stress and body forces. The remaining ones, however, are entirely interface related contributions, since $\nabla\chia$ and $\nabla\phia$ are non-zero only in the interface domain. For instance, $\langle \Ph\nabla\chia \rangle$ samples the fluid pressure in the interface domain only. Moreover, both $\chia$ and $\phia$ take values in the interval $[0,1]$ and vary over the same interface width so that their gradients are roughly of the same order of magnitude. Therefore, it is reasonable to define an interface-pressure, $\PhaI$, via
\begin{equation}
\langle \Ph \nabla \chia \rangle    = \PhaI \nabla \phia.
\label{eq:def-PaI}
\end{equation}
Similarly, we define an interface-stress tensor, $\tauaI$, 
\begin{equation}
\langle \vec \tau \cdot \nabla \chia \rangle    = \tauaI \cdot \nabla \phia.
\label{eq:def-tauaI}
\end{equation}
With these definitions, the pressure related term in Eq.~(\ref{eq:DI-version-of-NS-equation-rhs-1}) becomes
\begin{equation}
-\Pha\nabla\phia + \langle \Ph \nabla \chia \rangle  = (\PhaI -\Pha) \nabla \phia.
\label{eq:interface-pressure-term-in-NS-equation-1}
\end{equation}
Since spatial pressure variations homogenize with the speed of sound, we expect the value of pressure close to the interface to be essentially equal to the bulk pressure in the low Mach number flows considered in the present work. Therefore, we can neglect this pressure difference and set $\PhaI-\Pha=0$ in the momentum balance equation. As to the stress tensor, we have
\begin{equation}
\vec \taua \cdot \nabla \phia - \langle \vec \tau \cdot \nabla \chia \rangle  = (\taua -\tauaI)\cdot \nabla \phia.
\label{eq:interface-stress-term-in-NS-equation-1}
\end{equation}
In contrast to pressure, which couples to density variations and thus homogenizes with the speed of sound, stress forces arise from velocity gradients in the fluid and are particularly important close to the interface, where such gradients are usually high (despite the fact that fluid velocity itself can be fairly small). Therefore, we keep the interface stress term in the Navier-Stokes equation. Accounting for these considerations, and gathering the above results together, the volume-averaged version of the Navier-Stokes equation becomes
\begin{align}
&\rhoa (\partial_t \ua + \ua\cdot\nabla\ua) = -\nabla \Pha +\nabla\cdot\taua +\Fba + (\taua-\tauaI)\cdot \frac{\nabla\phia}{\phia} .
\label{eq:volume-averaged-NS-equation-3}
\end{align}
where we have also divided both sides of Eq.~(\ref{eq:volume-averaged-NS-equation-3}) by $\phia$. Equation~(\ref{eq:volume-averaged-NS-equation-3}) is nothing but the momentum balance equation for a fluid of phase $\alpha$ experiencing a drag force of the form
\begin{equation}
{\Fdrag}_\alpha=(\taua-\tauaI)\cdot \frac{\nabla\phia}{\phia}.
\label{eq:Fdrag-is-stress-difference-at-interface}
\end{equation}
at the interface. With this, the diffuse-interface version of the momentum balance equation takes its final form
\begin{align}
&\rhoa (\partial_t \ua + \ua\cdot\nabla\ua) = -\nabla \Pha +\nabla\cdot\taua +\Fba + {\Fdrag}_\alpha .
\label{eq:volume-averaged-NS-equation-4}
\end{align}
Calculation of $\Fdrag$ from a first-principle study requires a detailed knowledge of the fluid-solid interaction forces and the structure of the interface on the atomic scale. Such a calculation is beyond the scope of the present study. Fortunately, such detailed knowledge is not necessary. Rather, it is sufficient to make sure that the drag force hinders the fluid motion relative to the solid body such that the no-slip boundary condition is satisfied. \blue{Therefore, we follow previous studies and adopt an empirical model which was first suggested in~\cite{Beckermann1999} and successfully tested also in~\cite{Subhedar2015},
\begin{equation}
    {\Fdrag}_\alpha=-A_u \; \mu_\alpha\,\frac{\phia\,\phib^2}{\eta^2}(\ua-\ub),
\label{eq:Fdrag-empirical-law}
\end{equation}
where $A_u>0$ is a coupling constant. This model ensures that the drag force is proportional to the solid-fluid velocity difference, $\Fdrag \propto  (\ua-\ub)$, where $\alpha$ and $\beta$ stand for fluid and solid phases.}
The dependence on the phase field functions $\phia$ and $\phib$ is chosen such that the interface term vanishes in the bulk domains, where either $\phia=1$ (which implies $\phib=0$) or $\phib=1$ ($\to \phia=0$). 

The parameter $A_u$ can be tuned to make sure that results obtained within the  diffuse interface formalism are as close as possible to the sharp interface limit~\cite{Subhedar2015}. In the case of incompressible Navier-Stokes equation, the optimum value of $A_u$ was determined in Ref.~\cite{Subhedar2015} (where it was denoted as $h$. Here we use $A_u$, since the symbol $h$ is used to denote enthalpy, see Eq.~(\ref{eq:energy-cons})). In the present problem of reactive compressible fluid flow, the best choice for $A_u$ will be determined in the benchmark section.

\subsubsection{Species equations at a diffuse interface}
\label{sec:Species-equations-at-a-diffuse-interface}

We now apply the averaging procedure to Eq.~(\ref{eq:species-cons}) and obtain,
\begin{align}
\langle \chi_{\alpha} \partial_t (\rho Y_k)\rangle  + \langle \chi_{\alpha}  \nabla
\cdot(\rho Y_k \bm{u} )  \rangle &= -\langle \chi_{\alpha}  \nabla\cdot\bm{j}_k \rangle  + \langle \chi_{\alpha} \dot{\omega}_k  \rangle  
\label{eq:volume-averaged-species-equation-1}.
\end{align}
It is useful to note the similarity between the terms on the left hand side of Eq.~(\ref{eq:volume-averaged-species-equation-1}) with those on the lhs of  Eq.~(\ref{eq:vol-average-over-chia-x-mass-conservation-equation}). Indeed, the former can be obtained from the latter via replacing $\rho$ by $\rho Y_k$. Using this similarity, we transfer the lhs of Eq.~(\ref{eq:volume-averaged-continuity-equation})
---which is a reformulated version of the lhs of Eq.~(\ref{eq:vol-average-over-chia-x-mass-conservation-equation})--- to the present case and write
\begin{align}
\text{lhs of Eq.~(\ref{eq:volume-averaged-species-equation-1})}&=\partial_t (\phia \rhoa \Yka)  +   \nabla \cdot(\phia\rhoa\Yka\ua) \nonumber\\
&- \langle \rho Y_k (\vec u-\vI) \cdot\nabla \chia\rangle.
\label{eq:volume-averaged-species-equation-1-lhs}
\end{align}
To rewrite the first term on the rhs of Eq.~(\ref{eq:volume-averaged-species-equation-1}), we use 
Eq.~(\ref{eq:average-of-chi-nablafvec}) and make the identification $\vec f \equiv \vec j_k$. This yields 
\begin{align}
\langle \chi_{\alpha}  \nabla\cdot\bm{j}_k \rangle &= \nabla \cdot (\phia\,\vec j_{k,\alpha}) - \langle \vec j_k \cdot  \nabla \chia \rangle.
\label{eq:chia-nabla-jk}
\end{align}
The last term on the rhs of Eq.~(\ref{eq:volume-averaged-species-equation-1}) is readily given by (see Eq.~(\ref{eq:def-volume-average-fchia})),
\begin{align}
 \langle \chi_{\alpha} \dot{\omega}_k  \rangle  &= \phia \dot{\omega}_{k,\alpha}.
\label{eq:chia-dot-omegak}
\end{align}
Inserting equations~(\ref{eq:volume-averaged-species-equation-1-lhs})-(\ref{eq:chia-dot-omegak}) into Eq.~(\ref{eq:volume-averaged-species-equation-1}) and rearranging terms yields,
\begin{align}
&\partial_t (\phia \rhoa \Yka)  +   \nabla \cdot(\phia\rhoa\Yka\ua)- \langle \rho Y_k (\vec u-\vI) \cdot\nabla \chia\rangle \nonumber\\
&= - \nabla \cdot (\phia\,\vec j_{k,\alpha}) + \langle \vec j_k \cdot  \nabla \chia \rangle +\phia \dot{\omega}_{k,\alpha}.
\label{eq:volume-averaged-species-equation-2}
\end{align}
We apply the product rule of differentiation to Eq.~(\ref{eq:volume-averaged-species-equation-2}), rearrange terms which have $\Yka$ as a common factor and use Eq.~(\ref{eq:volume-averaged-continuity-equation}) to arrive at
\begin{align}
&\rhoa \phia \partial_t \Yka + \rhoa \phia \ua \cdot  \nabla \Yka + \Yka \langle \rho (\vec u-\vI)\cdot\nabla \chia \rangle - \langle \rho Y_k (\vec u - \vI) \cdot  \nabla \chia \rangle \nonumber  \\
&= - \phia  \nabla \cdot \jka - \jka \cdot  \nabla  \phia 
+\langle \jk \cdot  \nabla \chia \rangle + \phia\, \dot{\omega}_{k,\alpha} . 
\label{eq:volume-averaged-species-equation-3}
\end{align}
Adding the last two terms on the lhs of Eq.~(\ref{eq:volume-averaged-species-equation-3}) yields 
$\langle \rho(\Yka-Y_k)(\vec u-\vI)\cdot\nabla\chia \rangle$, where we used the fact that $\Yka$ is constant on the subgrid scale and can, therefore, be taken into the volume averaging integral.
This expression contains the product of fluctuations of composition and fluid velocity inside the interface and can, therefore, be neglected. To see the smallness of fluctuations, we recall that solid drag ensures that fluid velocity approaches the interface velocity so that $\vec u-\vI \ll \vec u$. Moreover, concentration variations homogenize quickly inside the thin interface region so that $\Yka-Y_k\ll Y_k$. 

As to the rhs of Eq.~(\ref{eq:volume-averaged-species-equation-3}), we introduce interface-related current density much in the same way as we did for interface-related pressure and stress tensor, Eq.~(\ref{eq:def-PaI}) and Eq.~(\ref{eq:def-tauaI}),
\begin{equation}
    \langle \jk \cdot \nabla \chia \rangle = \jkaI \cdot \nabla \phia 
\label{eq:def-jkaI}
\end{equation}
Neglecting the above discussed product of fluctuations and
using the defining Eq.~(\ref{eq:def-jkaI}), one obtains from Eq.~(\ref{eq:volume-averaged-species-equation-3}),
\begin{align}
\rhoa ( \partial_t \Yka  + \ua \cdot  \nabla \Yka ) &= - \nabla \cdot \jka 
- (\jka - \jkaI) \cdot  \frac{\nabla \phia} {\phia}
+ \dot{\omega}_{k,\alpha}, 
\label{eq:volume-averaged-species-equation-4}
\end{align}
where we have divided both sides of the equation by the common factor $\nabla \phia$. In similarity to the momentum balance equations, diffuse-interface version of species transport equation in a system composed of two phases has essentially the same shape as in the single-phase case with the only difference being encoded in an interface term, here the second term on the rhs of Eq.~(\ref{eq:volume-averaged-species-equation-4}). This term will play a key role for modelling solid combustion, where various reacting species are consumed/produced at the fluid-solid interface. In the present study, however, the solid walls are considered to be inert and reactions are restricted to the fluid phase. Therefore, and due to the impermeable nature of the walls, there is no flux of chemical species in the direction normal to the interface. This condition is equivalent to setting $(\jka - \jkaI) \cdot  \frac{\nabla \phia} {\phia}=0$ in Eq.~(\ref{eq:volume-averaged-species-equation-4}). One thus obtains,
\begin{align}
\rhoa ( \partial_t \Yka  + \ua \cdot  \nabla \Yka ) &= - \nabla \cdot \jka + \dot{\omega}_{k,\alpha}.
\label{eq:volume-averaged-species-equation-5}
\end{align}
In order to calculate the mass reaction rate of species $k$, the following equation is used \cite{Poinsot2005},
\begin{align}
\dot{\omega}_\text{k} =& M_k \sum_{j=1}^{M}  s_{kj} Q_j,  \label{eq:8}
\end{align}
where $M$ is the number of chemical reactions which take place. $Q_j$ and $s_{kj}$ are the rate of progress of reaction $j$ and the difference of the molar stoichiometric coefficient of species $k$ in productions and reactants in reaction $j$, respectively~\cite{Poinsot2005}. 

\subsection{DI-version of the energy transport equation}
Building upon the previous subsection, we proceed to obtain the diffuse-interface version of the energy transport equation. For this purpose, we note the similarity of the structure of the energy transport (Eq.~(\ref{eq:energy-cons})) and species transport (Eq.~(\ref{eq:species-cons})) equations. Indeed, all but one of the terms on the rhs of Eq.~(\ref{eq:energy-cons}) can be obtained from Eq.~(\ref{eq:species-cons}) by the replacements $Y_k\to h$, $\jk \to \vec q$ and $\dot{\omega}_k \to \dot{Q}$. The only term which then needs a special treatment is $\partial_t P_0$ on the rhs of Eq.~(\ref{eq:energy-cons}). Multiplying this term with $\chia$ and volume averaging yields,
\begin{equation}
\langle  \chia  \partial_t P_0 \rangle = \langle  \chia  \partial_t \Pna \rangle = \langle \chia \rangle\,  \partial_t \Pna = \phia\, \partial_t \Pna, 
\label{eq:volume-averaged-chia-partial-t-Pnull}
\end{equation}
where we used the fact that $P_0$ is homogeneous and can thus be taken out of the volume average integral. 

Using the above mentioned analogy between the energy balance and species transport equations, we perform the corresponding replacements on Eq.~(\ref{eq:volume-averaged-species-equation-4}) and add the volume averaged pressure term (rhs of Eq.~(\ref{eq:volume-averaged-chia-partial-t-Pnull})) to its right hand side and thus obtain the volume-averaged energy balance equation,
\begin{align}
\rhoa ( \partial_t \ha  + \ua \cdot  \nabla \ha ) &= - \nabla \cdot \qa 
- (\qa - \qaI) \cdot  \frac{\nabla \phia} {\phia}
+ \dot{Q}_{\alpha} +  \partial_t \Pna.
\label{eq:volume-averaged-enthalpy-equation-5}
\end{align}
As in the case of momentum balance and species mass transport equations, the multi-phase version of the transport equation for energy, Eq.~(\ref{eq:volume-averaged-enthalpy-equation-5}), has a form similar to the simple phase one, Eq.~(\ref{eq:species-cons}), with the effect of other phases appearing in an interface term, $\SIa \equiv -(\qa - \qaI) \cdot \nabla \ln(\phia)$. The significance of this source term, $\SIa$, depends on the boundary condition. When a uniform heat flux at the interface is utilized, this term vanishes. However, in the case of a uniform surface (or wall) temperature, heat is exchanged between the fluid and the wall in order to keep temperature at the prescribed value. A first-principle computation of the associated heat flux is beyond the scope of the present work. Therefore, we use the following empirical relation,
\begin{equation}
\SIa = A_T\, \lambdaa \frac{\phia\,\phib^2}{\eta^2} (\Ta -\Tb).
\label{eq:diffusion-heat-flux-3}
\end{equation}
In CFD applications, it is more convenient to express the heat equation in terms of temperature rather than enthalpy as defined in Eq.~(\ref{eq:energy-cons})~\cite{Poinsot2005}. To do so, we introduce the so-called sensible enthalpy via $h_s=\int_{T_0}^T c_{p}\;  dT$, where $c_p$ is heat capacity at constant pressure. With this, enthalpy can be written as $h=h_s + h_0$, where $h_0=h(T_0)$ (note that $h_s(T_0)=0$). Using $h_s$, we reformulate Eq.~(\ref{eq:energy-cons}) in terms of sensible enthalpy. For a fluid mixture with $N$ species, the reference enthalpy is given by $h_0=-\sum_{k=1}^{N} Y_k \Delta h^{o}_{f,k}$ where $\Delta h^{o}_{f,k}$ represents the formation enthalpy of species $k$ at a reference temperature of $T_0=298.15$ K. By incorporating these definitions and Eq.~(\ref{eq:species-cons}) for species transport, after substituting $h_s$ into Eq.~(\ref{eq:energy-cons}), we obtain the following expression~\cite{Poinsot2005},
\begin{equation}
   \rhoa c_p (\partial_t \Ta + \ua \cdot \nabla \Ta)  = \nabla \cdot (\lambda_\alpha \nabla \Ta) + \partial_t \Pna + \nabla \cdot \Ta (\sum_{k=1}^{N} c_{p,k} \jka) + \dot{\omega}_{T,\alpha} + \SIa.
\label{eq:energy-transport-DI-version}
\end{equation}
where $\dot{\omega}_\text{T}$ is heat production rate and can be calculated via~\cite{Poinsot2005},
\begin{align}
\dot{\omega}_\text{T} =& - \sum_{k=1}^{N} \Delta h^{o}_{f,k} \dot{\omega}_\text{k}, \label{eq:9}
\end{align}
In the above equations, the viscous heating term in the temperature equation, and the Soret effect (mass diffusion induced by temperature gradient) are neglected since their contributions are not significant comparing to other terms. 

\blue{It is worth mentioning that, in the asymptotic limit of zero interface thickness ($\eta \to 0$), all the DI-equations given in this section converge to their corresponding sharp-interface counterparts. Results of such asymptotic analysis have been published for the case of phase-field equations combined with fluid dynamic equations (see, e.g., \cite{Subhedar2020}). An asymptotic analysis of the present set of equations runs along similar lines and is omitted here.}

\blue{For numerical integration of the above-described DI-version of transport equations for momentum, mass and energy, we use a} hybrid Lattice Bolzmann-Finite Difference (LB-FD) approach under low Mach number conditions. In this framework, the flow dynamics is modeled via a modified lattice Boltzmann method proposed in~\cite{Hosseini2019}. The energy and mass exchanges are computed using a FD scheme. This LB solver considers solely the compressibility arising from thermal variations. The following distribution function and source terms are employed to determine the velocity and pressure of the flow~\cite{Hosseini2020PhDTheis,Hosseini2019,Hosseini2020}:
\begin{align}
	\frac{\partial g_{\alpha}}{\partial t} &+ \bm{c}_\alpha \cdot \bm{\nabla}g_\alpha = -\frac{1}{\tau}(g_\alpha-g_\alpha^{\rm eq}) + \Xi_\alpha, \label{eq:pop-g} \\
	\Xi_\alpha &= (\bm{c}_\alpha - \bm{u}) \cdot [\bm{\nabla} {\rho \cs^2} (f^{\rm eq}/\rho - w_\alpha) + \Fb f^{\rm eq}/\rho] + 
	{\rho \cs^2}w_\alpha \bm{\nabla} \cdot \bm{u}  \label{eq:body-force},
\end{align}
where:
\begin{equation}
	f_\alpha^{\rm eq} = \rho w_\alpha\left(1+\frac{1}{\cs^2}\bm{c}_\alpha\cdot\bm{u}+\frac{1}{2\cs^4}{(\bm{c}_\alpha\cdot\bm{u})}^2 - \frac{1}{2\cs^2}\bm{u}^2\right) \label{eq:eq-pop-f},
\end{equation}
and
\begin{equation}
	g_\alpha^{\rm eq} = \cs^2 f_\alpha^{\rm eq} + w_\alpha (\Ph-\rho \cs^2) \label{eq:eq-pop-g}.
\end{equation}
$f_\alpha$, $w_\alpha$, $\tau$, $\cs $ and $\bm{c}_\alpha$ are conventional discrete populations, weights, relaxation coefficient, speed of sound and the lattice velocity along the lattice direction $\alpha$, respectively. The dynamic viscosity can be calculated by $\mu=\rho \cs^2(\tau_r-0.5)$. In the FD approach, the convective and diffusion terms are discretised using upwind and central schemes, respectively. The temporal terms are approximated with a first-order Euler discretisation. More detail of this methodology can be found in ~\cite{Hosseini2019}.

\section{Benchmark tests and validation}\label{sec:benchmark-and-validation}
This section presents results of numerical simulations, which serve to validate the above derived transport equations within a phase-field based approach. Along this line, we first test the momentum and heat exchange equations in the absence of reactions by considering the flow of air. The dynamic viscosity of air is computed using the Sutherland's law,
\begin{equation}
	\mu(T) = \mu^* \Big(\frac{T}{T^*}\Big)^{\frac{3}{2}}\;\frac{T^*+S}{T+S} \label{eq:muT},
\end{equation}
with $T^*=273\,$~K, $S=110.5\,$~K and $\mu^*= 1.68\times 10^{-5}$~Pa$\cdot$s for air. Since the Prandtl number is assumed to be constant in this study, thermal conductivity is calculated from,  
\begin{equation}
	\lambda(T) = \frac{\mu(T)c_p}{\Pra} \label{eq:lambdaT}.
\end{equation}

As a result of coupling to a diffuse fluid-solid interface, the model contains two new coupling parameters, $A_u$ and $A_T$ (see Eqs.~(\ref{eq:Fdrag-empirical-law}) and (\ref{eq:diffusion-heat-flux-3})) whose optimal choice is addressed here. To this end, simple test cases are selected and the results obtained within the phase field based diffuse interface approach are compared to their sharp interface counterparts. The sharp interface solution is obtained by switching off \label{page:SI-simulation-method} the coupling to the phase-field function in equations~(\ref{eq:volume-averaged-NS-equation-4}), (\ref{eq:volume-averaged-species-equation-5}) and (\ref{eq:energy-transport-DI-version}).

\begin{figure}[t]
	\centering
    \hideimage{\includegraphics[width=0.9\linewidth]{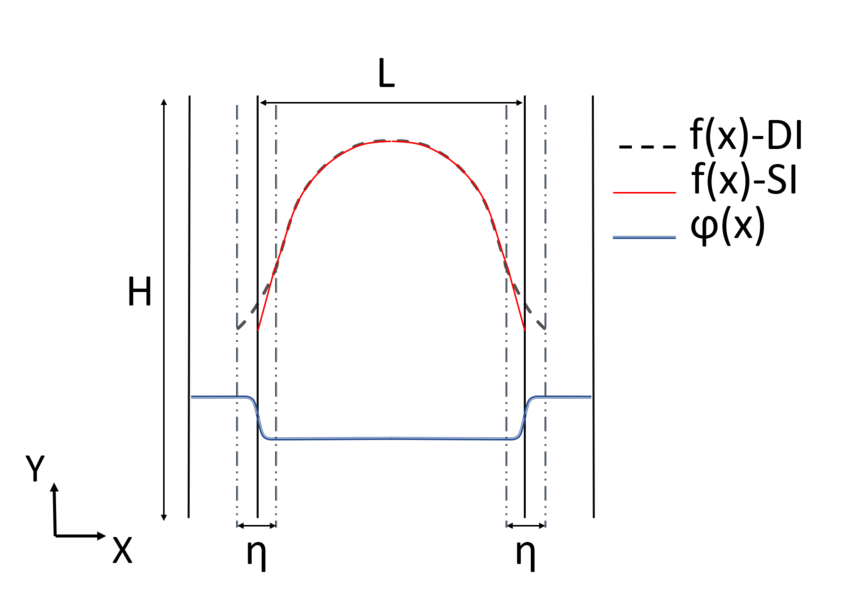}}
    \caption{A schematic view of the expected qualitative behavior of a quantity $f$ within SI and DI methods. Since the main difference between sharp and diffuse interface formulations is in the interface region (the underlying equations being identical in the bulk-fluid domain), one expects the SI and DI versions of $f$ to be different there. At the same time, it is an essential requirement that results obtained with the DI-formulation be as close as possible to the SI-solution outside the interface. The function $\varphi(x)$ depicts the solid fraction, which varies between 1 (inside the walls) and 0 (in the fluid domain). The parameter $L$ is the distance between the two solid walls, $H$ denotes the inlet-outlet distance and $\eta$ is the thickness of the diffuse fluid-solid interface.}
	\label{fig:SI-vs-DI}
\end{figure}

\blue{In this study, the variation of $\varphi$ as a function of the distance from the exact position of the solid wall is defined as,}
\begin{equation}
    \blue{\varphi(x) = \frac{1}{2} \left[ 1 - \sin\left( \frac{\pi x}{\eta} \right) \right],} \label{eq:def-phi}
\end{equation}
where $x$ denotes the distance from the wall and $\eta$ is the interface thickness. An illustration of the simulation domain is shown in Fig.~\ref{fig:SI-vs-DI}. The figure also contains a schematic drawing of the expected qualitative behavior of a quantity, $f$, within DI and SI methods.
Since sharp and diffuse interface formulations of fluid flow and heat transfer equations differ only in the fluid-solid transition zone, one expects the SI and DI methods to yield different results in this domain but be close to each other in the bulk-fluid region, where the set of equations for the two formulations become identical. The discrepancy between the DI and SI approaches inside the interface does not pose a problem, since the behavior of the physical quantities in this domain is not of primary interest\footnote{In the opposite case, i.e., if one is interested in interface properties, the present methodology can still be applied. For this purpose, a thorough study is necessary to derive the fluid-solid coupling term from basic physical principles.}. In the bulk fluid domain, on the other hand, one is interested in obtaining the same results regardless of the underlying numerical scheme.

To put the comparison of the data obtained within the DI and SI methods for a property $f$ on a quantitative ground, we calculate the mean standard error of the diffuse interface version of this quantity, $f_\text{DI}$ with respect to the sharp interface solution, $f_\text{SI}$,
\begin{equation}
	\text{MSE} = \frac{1}{L-\eta}\int_{\eta/2}^{L-\eta/2} \Big[\frac{f_\text{SI}(x)-f_\text{DI}(x)}{f_\text{SI,max}}\Big]^2\;d x
	\label{eq:mean-standard-error-f},
\end{equation}
where $L$ is the channel width and $\eta$ is the interface thickness. In line with the above comments, the integration range in Eq.~(\ref{eq:mean-standard-error-f}) is restricted to the bulk fluid domain, where agreement between the two approaches is expected.

\subsection{Test of the isothermal flow}
\label{sec:Test-of-isothermal-flow}
We first analyze the behavior of the DI-model for the simpler case of fluid flow in the absence of heat transfer, Eq.~(\ref{eq:volume-averaged-NS-equation-4}). A similar analysis has been conducted in one of our previous works for a gravity driven flow with inlet/outlet periodic boundary conditions~\cite{Subhedar2015}. In the present work, we consider a channel with open boundaries. In contrast to the periodic boundary condition, whose implementation is rather straightforward and poses no additional constraints on numerical accuracy, a sophisticated non-reflecting scheme~\cite{poinsot1992} is necessary to avoid numerical instabilities due to reflections of pressure/sound waves at an open boundary. The present setup thus necessitates a separate study on its own right in order to tune the parameter, $A_u$, which appears in the diffuse interface-coupling term, Eq.~(\ref{eq:Fdrag-empirical-law}), in such a way that the best agreement with the SI-solution is obtained. The temperature of the walls and the fluid are set to the same value (both at $300$ K) so that heat transfer does not play a role.

\begin{figure}[t]
	\centering
    \hideimage{\includegraphics[width=1.0\linewidth]{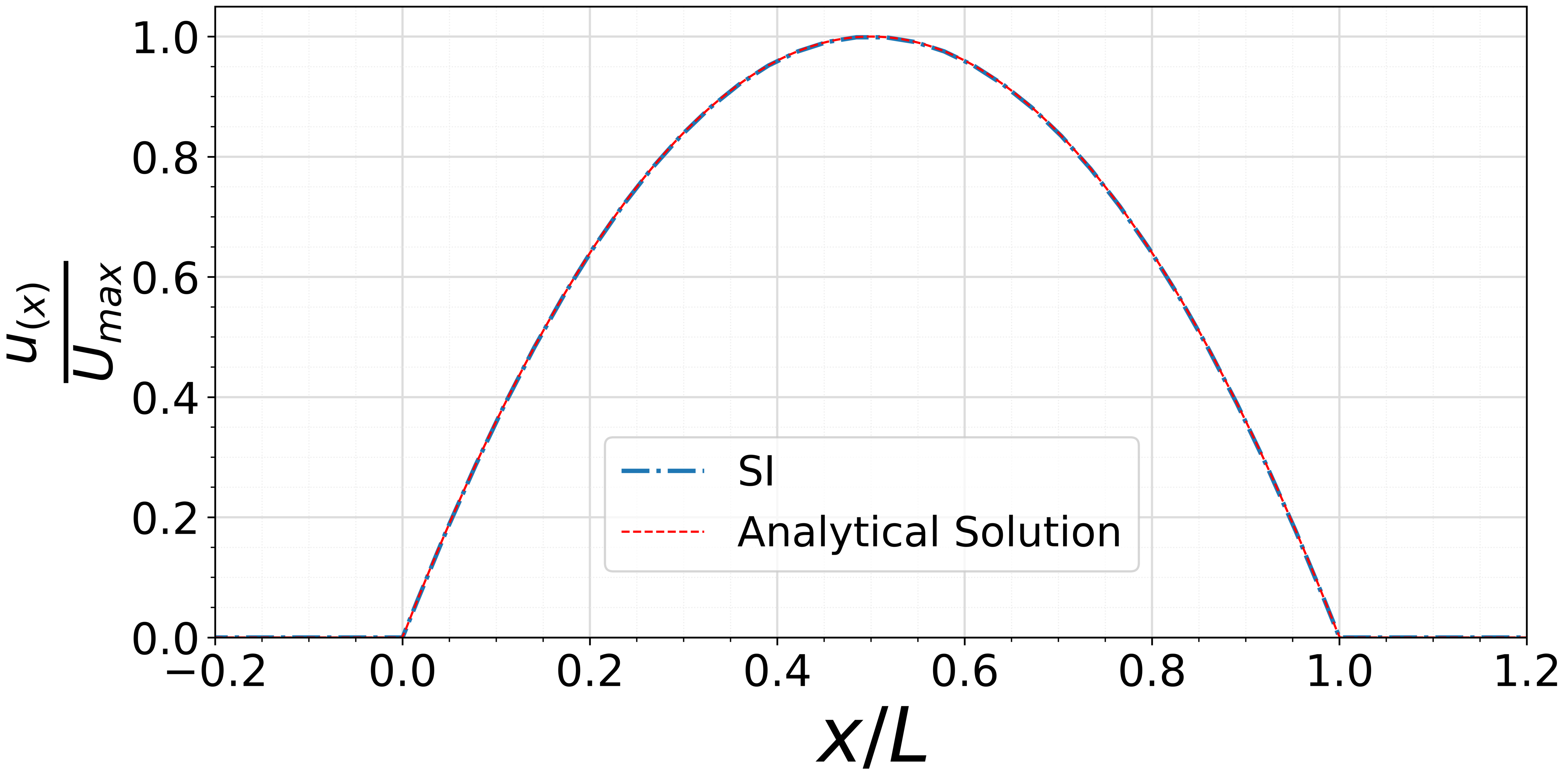}}
    \caption {Comparison of sharp interface simulation results with the analytical solution for Poiseuille flow at $Re=100$ with a channel width of $40 \Delta x$. The simulation accurately captures the analytical velocity profile, demonstrating the validity of the numerical approach.}
	\label{fig:SI-vs-Analytical}
\end{figure}
\begin{figure}[t]
	\centering
    \hideimage{\includegraphics[width=1.0\linewidth]{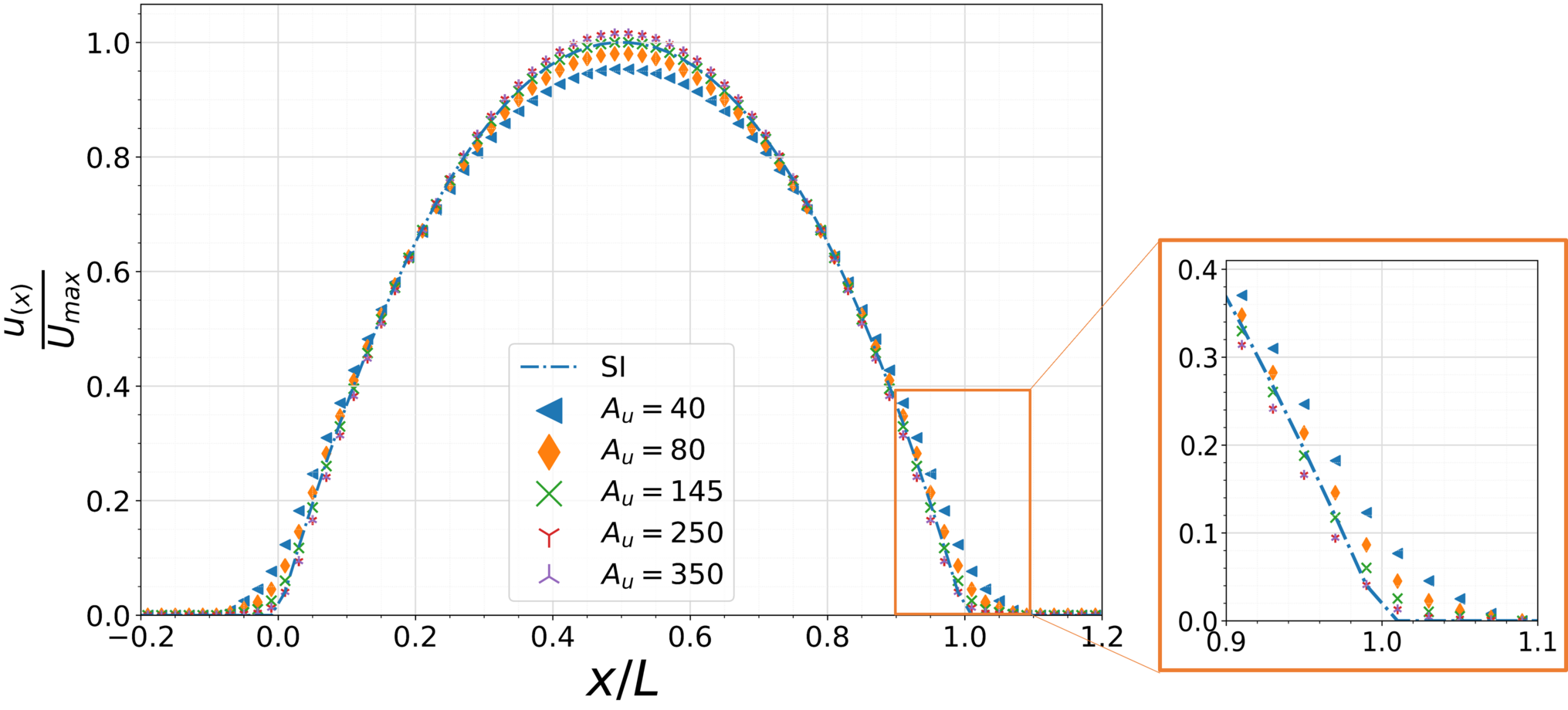}}
    \caption{Velocity profiles obtained within the phase-field based DI-method and the corresponding SI-solution at a Reynolds number of $\Rey=100$. Each DI-curve corresponds to a choice of $A_u$ as indicated in the curve-legend. The image on the right is a zoom into the interface region. For $A_u=145$, a very good agreement between the DI and SI-calculations is observed. The channel width is $L=50\Delta x$, the interface thickness is $\eta=8\Delta x$ and the inlet-outlet distance is chosen to be $H=5L$. The horizontal axis covers the range $x\in [-\frac{\eta}{2},\;  L+\frac{\eta}{2}]$. We thus have $x\in [-\frac{\eta}{2L},\;  1+\frac{\eta}{2L}]$. In the limit of a sharp interface, $\eta/L\to 0$, the investigated range becomes identical to [0, 1].}
	\label{fig:DI-vs-SI-diff_Au}
\end{figure}

To evaluate the DI results by comparing them with SI results, we first validate our solver for sharp interface equations. For this purpose, we select a Poiseuille flow configuration, where a fluid enters a channel between two parallel planar walls with a uniform velocity profile at the inlet $u(x,y=0)=u_0$. For distances sufficiently far from the inlet, one expects a parabolic velocity profile to establish. To test the solver, we hence performed a simulation with a channel height significantly larger than its width ($H \geq 40L$). Results of these simulations are shown in Fig.~\ref{fig:SI-vs-Analytical}. As observed, the simulation accurately reproduces the expected analytical solution. In this setup, the channel width is $40 \Delta x$, and the Reynolds number is 100. It is important to note that in sharp interface simulations, bulk phases are separated at $\eta = 0.5$. Once the SI results were validated, we opted for smaller simulation domains in subsequent simulations to reduce computational cost. To address the effect of the coupling parameter $A_u$ on the flow, a number of simulations are performed with a uniform velocity profile at the inlet (i.e., $u(x)=\uinlet$ at $y=0$), and an open boundary at the outlet. Figure~\ref{fig:DI-vs-SI-diff_Au} depicts the thus obtained DI-results for five different values of $A_u$ as indicated in the legends. The sharp interface solution for the same setup is also shown in the plot. As seen from this comparison, the parameter $A_u$ influences the agreement between the phase-field based calculations and the reference SI-solution: Starting from $A_u=40$, the quality of the simulated data in the fluid-domain (i.e., beyond the interfaces) first improves for larger $A_u$ but then deteriorates for $A_u>145$.

\begin{figure}[t]
	\centering
    \hideimage{\includegraphics[width=1.0 \linewidth]{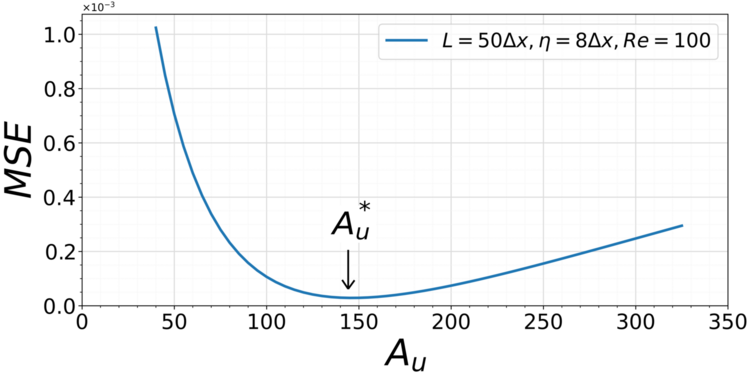}}
    \caption{The mean standard error, MSE in fluid velocity (evaluated at a section at $y=0.9H$) versus the coupling parameter, $A_u$ (Eq.~(\ref{eq:Fdrag-empirical-law})). In the present setup, the fluid is confined between two parallel planar solid walls and a uniform inlet velocity is imposed (The outlet is considered to be open). The data reveal the existence of a minimum in MSE as a function of $A_u$. The position of this minimum defines the optimal coupling parameter, $\Austar$. The Reynolds number is $\Rey=100$. The channel width is $L=50\Delta x$, the interface thickness is $\eta=8\Delta x$ and the inlet-outlet distance is chosen to be $H=5L$.}
	\label{fig:A_u-vs-MSE}
\end{figure}

\begin{figure}[t]
	\centering
\hideimage{\includegraphics[width=1.0 \linewidth]{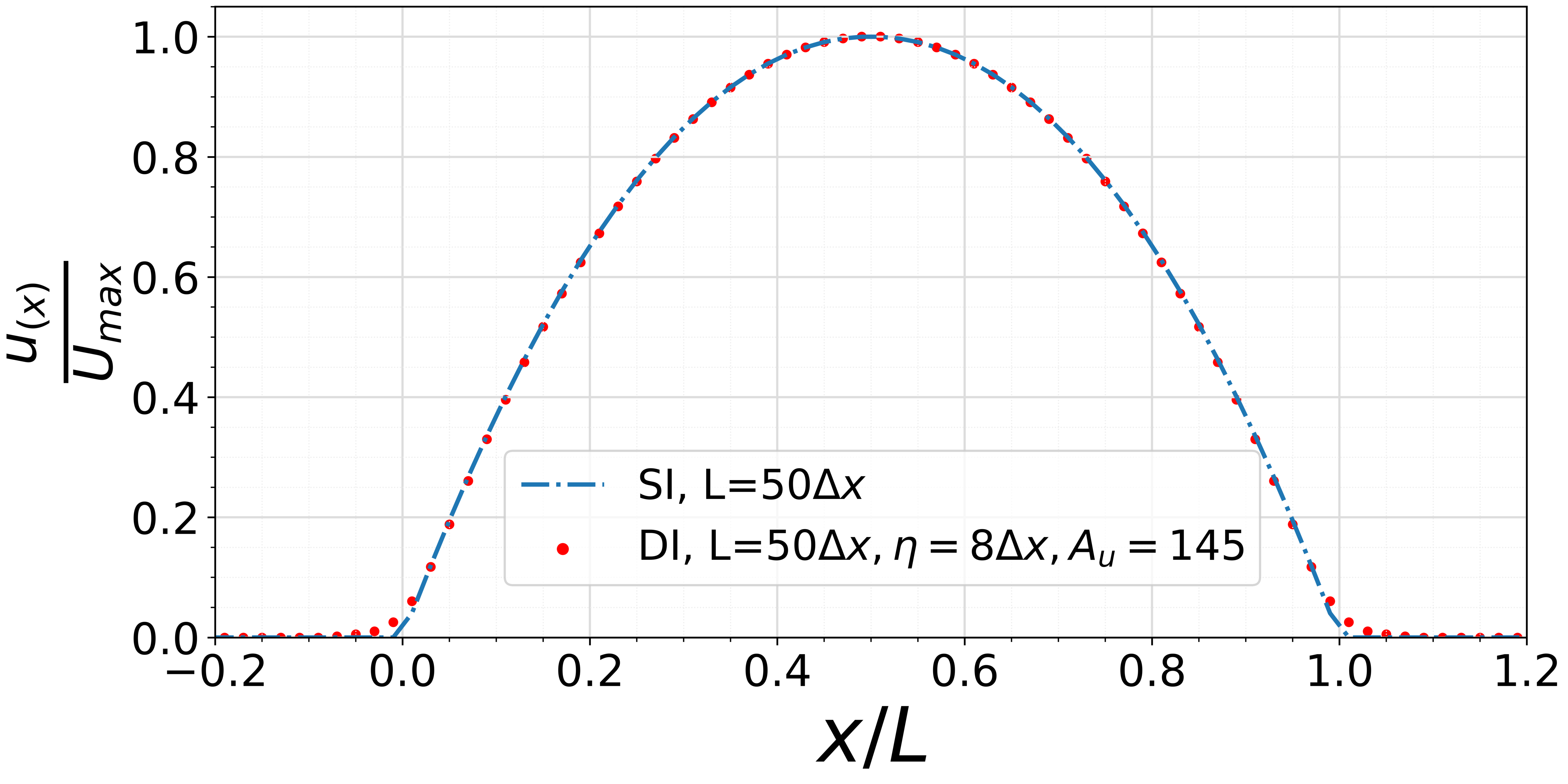}}
    \caption{The normalized velocity profile across the cross section at $y=0.9H$ versus the normalized transverse coordinate, $x/L$ at the optimal value of the coupling parameter, $\Austar=145$. The symbols give the DI-data and the dashed-dotted line the SI-counterpart. Here, $L=50\Delta x$ is the channel width (wall-to-wall distance) and $U_\text{max}$ is the maximum flow velocity at the considered cross section.}
	\label{fig:DI-vs-SI-u}
\end{figure}

In a systematic investigation of this issue, DI-based simulations at various $A_u$ are performed and the resulting velocity profiles are analyzed with regard to deviations from the sharp interface solution. Figure~\ref{fig:A_u-vs-MSE} shows the mean standard deviation of the simulated velocity profiles ($f\equiv u$ in Eq.~(\ref{eq:mean-standard-error-f})) versus $A_u$ for constant $L$ and $\eta$. Notably, a distinct point is observed where the deviation is minimized, represented by $A_u=\Austar$. The simulated flow field at this optimal coupling parameter is compared to the SI-solution in Fig.~\ref{fig:DI-vs-SI-u}. As expected, a very good agreement is observed between the both velocity profiles.

\begin{figure}[t]
	\centering
    \hideimage{\includegraphics[width=1.0\linewidth]{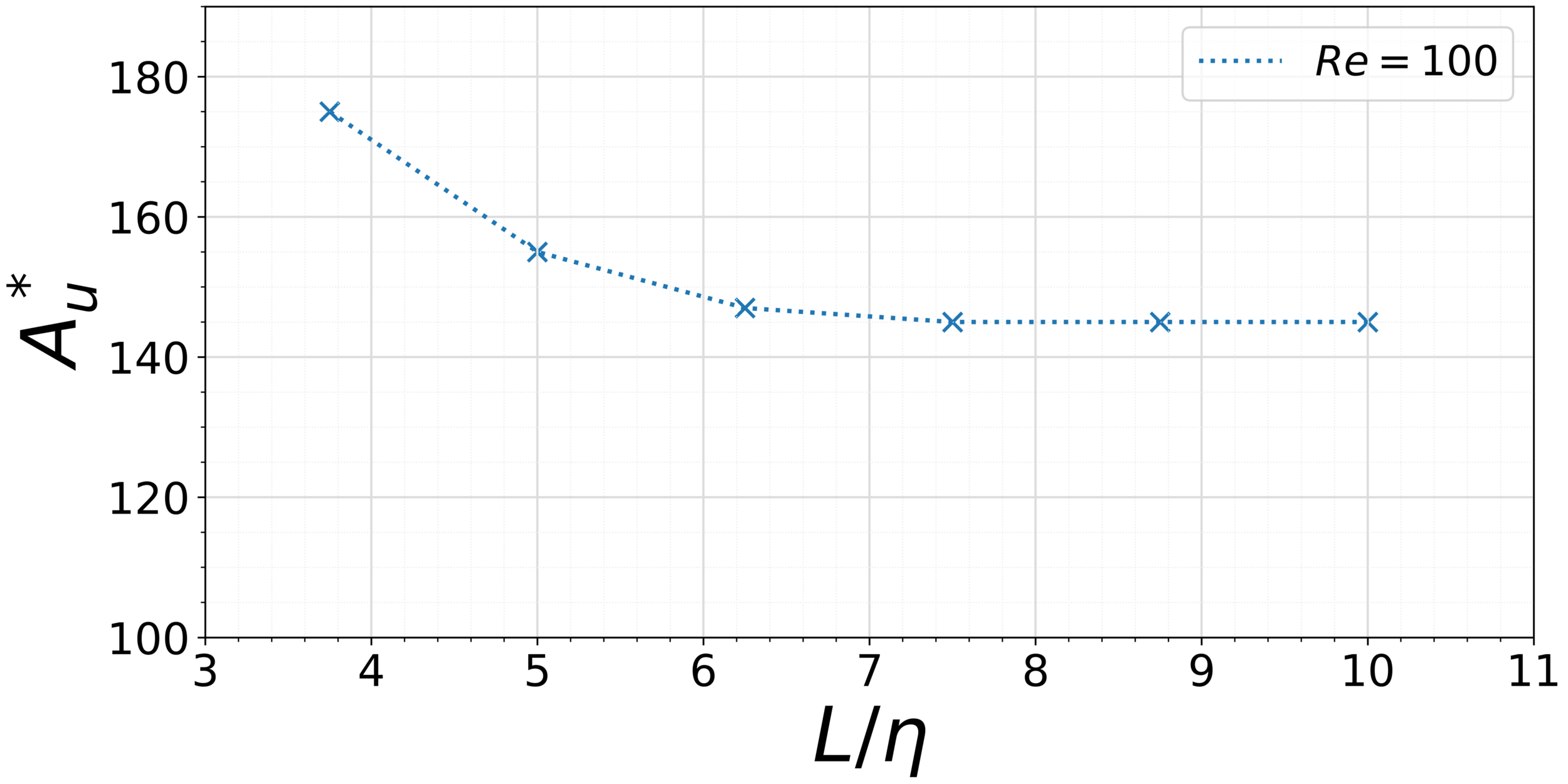}}
    \caption{Variation of the optimal coupling parameter for momentum exchange, $\Austar$, with the ratio of the system size to the interface thickness, $L/\eta$, for a constant Reynolds number of $\Rey=100$. Result for $\Austar$ becomes independent of interface thickness, provided that  $L \geq 6 \eta$.}
	\label{fig:Au*-vs-L/Eta}
\end{figure}

Since the phase field method introduces a numerical length scale --- the interface thickness $\eta$ --- into the problem, it is important to examine its possible influence on the optimal coupling parameter for fluid-solid momentum exchange, $\Austar$. On general grounds, one expects the numerical interface thickness to affect the simulation results if the channel width $L$ is not sufficiently large compared to $\eta$. To examine this issue, a series of simulations have been conducted and $\Austar$ has been determined for various system sizes, $L$, while keeping $\eta$ constant. This study amounts to sampling a two-dimensional parameter space: For each channel width $L$, a series of simulations are conducted for different values of $A_u$ in order to determine the optimal coupling constant $\Austar$. This procedure is then repeated for different values of $L$. To exclude possible effects of the Reynolds number, the fluid  velocity at the inlet is adjusted such that $\Rey$ remains constant. Results of these simulations are plotted in Fig.~\ref{fig:Au*-vs-L/Eta}. As expected, some effect of the numerical interface thickness is visible for small $L/\eta$. However, this influence decays quickly and becomes essentially negligible already for $L/\eta\ge 7$. This observation is encouraging and makes diffuse-interface simulations attractive as one does not need to choose exceedingly large system sizes for a reliable simulation of the problem.

\begin{figure}[t]
	\centering
    \hideimage{\includegraphics[width=1.0\linewidth]{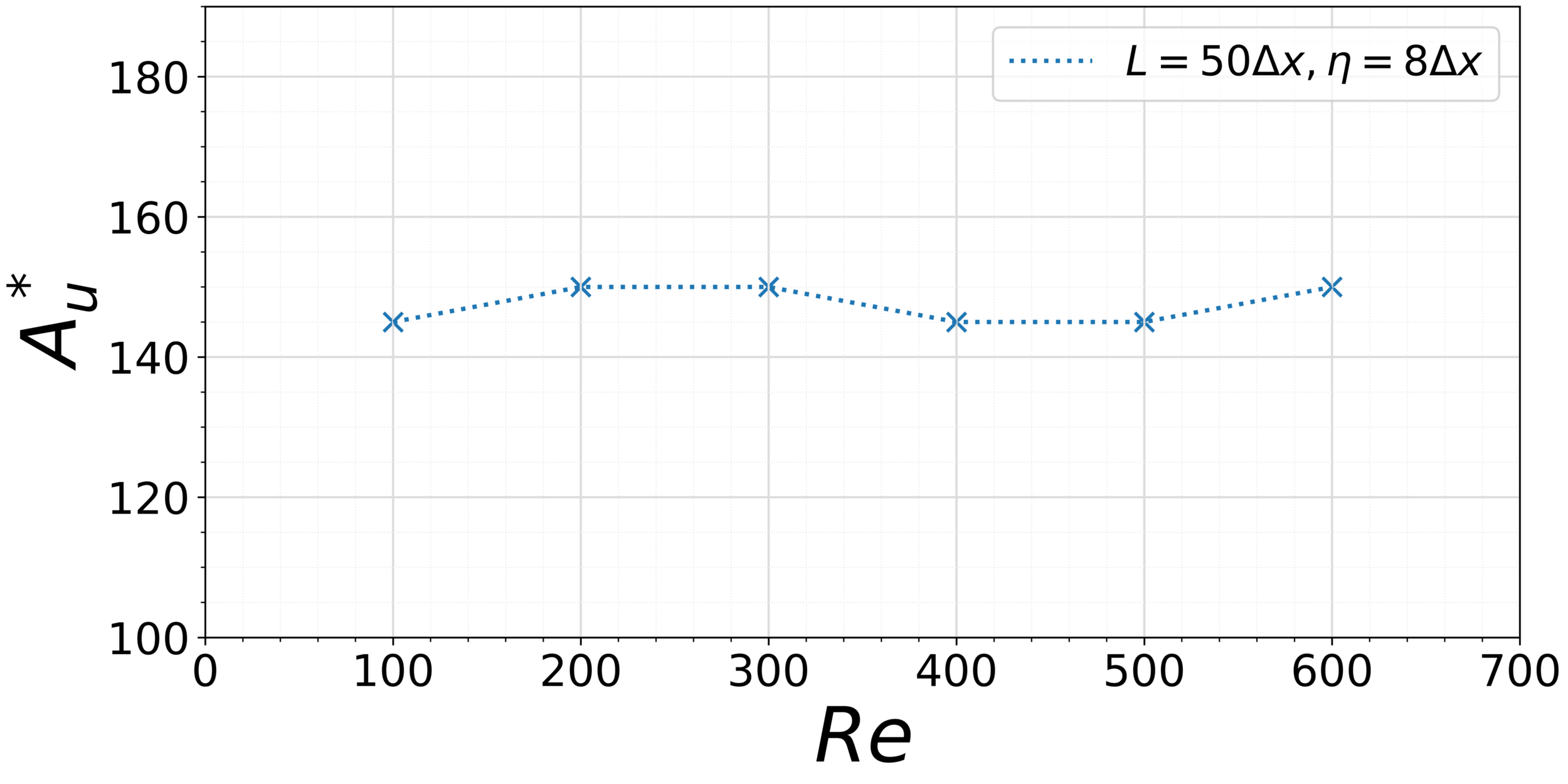}}
     \caption{The optimal coupling parameter for momentum exchange, $\Austar$, versus Reynolds number for a constant system size. Here, the Reynolds number is tuned in the range $\Rey=[ 100, 600]$ by a variation of the inlet flow velocity. From this data, one can conclude that flow velocity has little effect on $\Austar$.}
	\label{fig:Au*-vs-Re}
\end{figure}

Next, the possible effect of the Reynolds number on $\Austar$ is studied. To accomplish this task, $L$ and $\eta$ are kept constant and $\uinlet$ is varied such that the corresponding Reynolds numbers falls in the range $\Rey=100-600$. Again, a two-dimensional parameter space, now spanned by $\Rey$ and $A_u$, is sampled to obtain $\Austar$ as a function of $\Rey$. As shown in Fig.~\ref{fig:Au*-vs-Re}, $\Austar$ is rather insensitive to variations of the inlet fluid velocity for constant system size and interface thickness. This is an interesting observation as it supports the idea  that, for the velocities of interest to the present study, non-linear velocity terms in the fluid-solid momentum exchange are negligible so that a linear relation as given in Eq.~(\ref{eq:Fdrag-empirical-law}) is sufficient for the purpose of the present study.

It is instructive to compare the optimal value of $A_u$ in the present study ($\Austar=145$) with the results reported in~\cite{Subhedar2015} (note that this parameter is called $h^*$ in that reference). For this purpose, we have performed a simulation for $L=40\Delta x$, $\eta=6\Delta x$ and $\Rey=1$ corresponding to Fig.~5b in~\cite{Subhedar2015}. Within the present model, we obtain $\Austar=325$ instead of $h^*=241$. This difference can be traced back to --- and emphasizes the importance of --- boundary conditions and the details of the numerical schemes for the value of $\Austar$: In Ref.~\cite{Subhedar2015}, periodic boundary conditions have been used which require no special treatment, whereas the present study uses open boundaries with a non-reflecting numerical scheme at the outlet~\cite{poinsot1992} (see section~\ref{sec:Test-of-isothermal-flow}). These differences in boundary conditions and the numerical implementation have consequences on the contribution of the fluid-solid coupling term, Eq.~(\ref{eq:Fdrag-empirical-law}), to the momentum exchange at the diffuse interface. Tuning the coupling parameter $A_u$ serves to account for these influences in order to achieve the best agreement with the sharp interface solution.

\subsection{Non-isothermal flow}
\label{sec:Non-isothermal-flow}
The above validation of the DI-formulation of momentum balance equations sets the ground for a systematic study of DI-version of heat transport equations and the effects of the coupling parameter $A_T$, which controls the heat exchange at the fluid-solid interface, Eq.~(\ref{eq:diffusion-heat-flux-3}). Here, the same channel geometry is used as above but, instead of a uniform flow, a fully developed parabolic velocity profile is imposed at the inlet. A cold fluid enters from the inlet (at the bottom, $y=0$) into the domain between the two hot walls and is heated due to heat exchange with the walls as it passes along the channel. The temperature of both the walls is set to $\Twall=400$ K. The inlet fluid temperature is uniform and equal to $T(x,y=0)=\Tinlet=300$ K. The Reynolds number is $\Rey=100$. Moreover, $L$ and $\eta$ are kept constant at $L=50$ and $\eta=8$ grid spacing.

For this problem, a closed analytic solutions is not available. However, in the steady state (i.e., neglecting transient processes) and for a fully developed (i.e., $y$-independent) parabolic velocity profile, the heat transport equation (\ref{eq:energy-cons}) simplifies to $u(x) \partial T / \partial y =  \nu_T (\partial^2 T / \partial x^2 + \partial^2 T / \partial y^2)$, where $\nu_T$ is the heat diffusion coefficient (the walls are separated along the $x$-axis and  the fluid flows along the $y$-direction) This equation has the same mathematical structure  as the diffusion equation, where $y$ plays the sole of time and $nu_T/u(x)$ that of diffusion coefficient. Hence, inserting the parabolic velocity profile given in section~\ref{sec:Test-of-isothermal-flow} into this equation, it can be solved in an iterative way for the boundary conditions of interest to this work~\cite{incropera1996,Aydin2007}. This procedure is used to examine --- and gain confidence in --- the validity of the reference SI-solutions, which enters the evaluation of MSE for temperature, Eq.~(\ref{eq:mean-standard-error-f}).

\begin{figure}[t]
	\centering
    \hideimage{\includegraphics[width=0.9\linewidth]{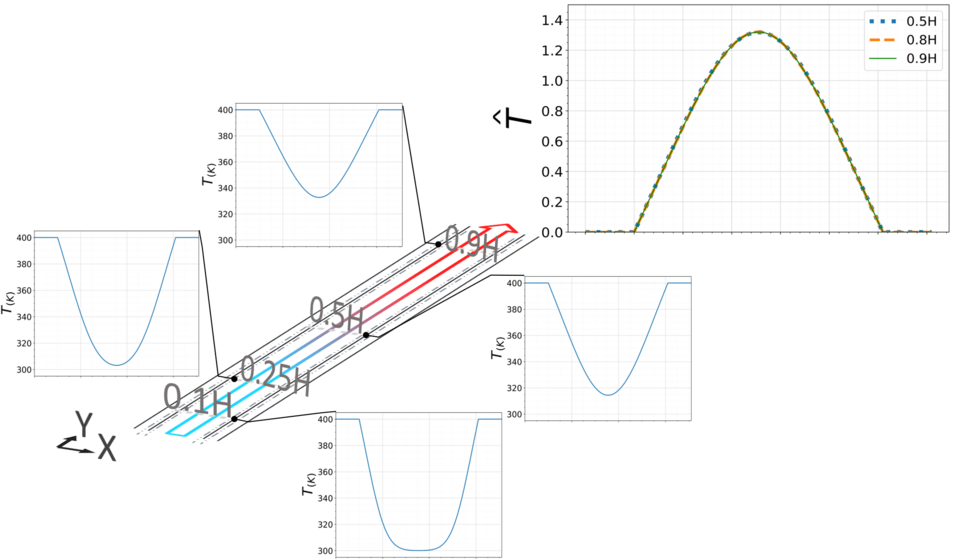}}
    \caption{A schematic view of the channel and the simulated temperature profiles at various cross sections, $y$, as indicated. Temperature profiles develop a similar shape for $y\ge 0.5H$. This is best seen on the upper right plot, where $\hat T(x,y)$ (Eq.~(\ref{eq:master-curve-for-T-profiles})) is shown versus $x$ for a number of cross sections.}
	\label{fig:T-profiles-w-channel-view}
\end{figure}

Figure~\ref{fig:T-profiles-w-channel-view} shows the simulated temperature profiles for a selected set of distances from the inlet. As expected, close to the inlet, the fluid temperature inside the channel is roughly uniform around the center and approaches the wall temperature in its proximity. Further away from the inlet, the heat arriving from the walls is felt to some extent also in the interior of the channel.

The typical distance, $l_T$, where wall temperature is felt inside the fluid, is given by the competition of convection and heat diffusion. A fluid parcel, which enters the channel at the inlet, needs a time of $t_u \sim y/\uc $ to travel the distance $y$, where $\uc$ is a characteristic flow velocity. During this time, heat diffusion reaches distances $l_T$, satisfying $l^2_T \sim \nu_T t_u \sim \nu_T y/\uc$. A fully developed temperature profile is established if $l_T=L/2$, i.e., if the heat from the walls reaches the center of the channel. The $y$-position, where this occurs defines the necessary length scale, $\HTfd$, for a fully developed temperature profile. Using this criterion, one obtains $H^\text{f.d.}_T \sim L^2 \uc /(4\nu_T)$. This scaling relation then yields $\HTfd\sim  \Rey \Pra L/4$, where we used the definitions of the Reynolds and Prandtl numbers, $\Rey=L \uc/\nu$ and $\Pra=\nu/\nu_T$, respectively. The constant of proportionality of this relation has been estimated in the literature with the result $\HTfd= 0.034\,\Rey\; \Pra\; L$~\cite{shah2003}. Since the Prandtl number is roughly constant with a value of $\Pra\approx 0.7$ in our simulations, the thermal entrance length obeys $\HTfd \approx 0.024\; \Rey\; L$. This length is less than the hydrodynamic entrance length, $\Hufd \approx 0.06 \Rey L$~\cite{white1990}. For a Reynolds number of $\Rey=100$, the above estimate yields an aspect ratio of $\HTfd/L=2.4$.

Results of these simulations are shown in Fig.~\ref{fig:T-profiles-w-channel-view}. It is seen from this plot that temperature profile evolves with distance $y$. However, it is possible to obtain a common shape by plotting the normalized difference of the local fluid and wall temperatures, which we define as
\begin{equation}
    \hat T (x,y) = \frac{\Twall-T(x,y)}{\Twall-\Tm(y)},
    \label{eq:master-curve-for-T-profiles}
\end{equation}
where, $\Tm$ is a cross-sectional mean temperature, defined via
\begin{equation}
\Tm(y)=\frac{\int_0^L \rho(x,y)\, u(x,y)\, c_p(T) T(x,y) dx}{\int_0^L \rho(x,y)\, u(x,y)\, c_p(T)dx}.
\label{eq:Tmean-def}
\end{equation}
The denominator in Eq.~(\ref{eq:Tmean-def}) gives the rate of energy transported by the flow per unit time and unit temperature across the cross section of the channel at the longitudinal position $y$. In the literature, it is often written as $\dot{m} c_p$~\cite{incropera1996}. 
As seen from Fig.~\ref{fig:T-profiles-w-channel-view}, $\hat T$ becomes independent from $y$ for $y\ge 0.5H$,
\begin{equation}
	\frac{\partial \hat T(x,y)}{\partial y}\approx 0  \quad (\text{for} \quad y \geq H/2).
	\label{eq:fully-developed-condition-for-T}
\end{equation}
Therefore, for the further analysis given below, we use the simulated temperature-data at a cross section of $y=0.9H$.
\begin{figure}[t]
	\centering
   \hideimage{ \includegraphics[width=1.0\linewidth]{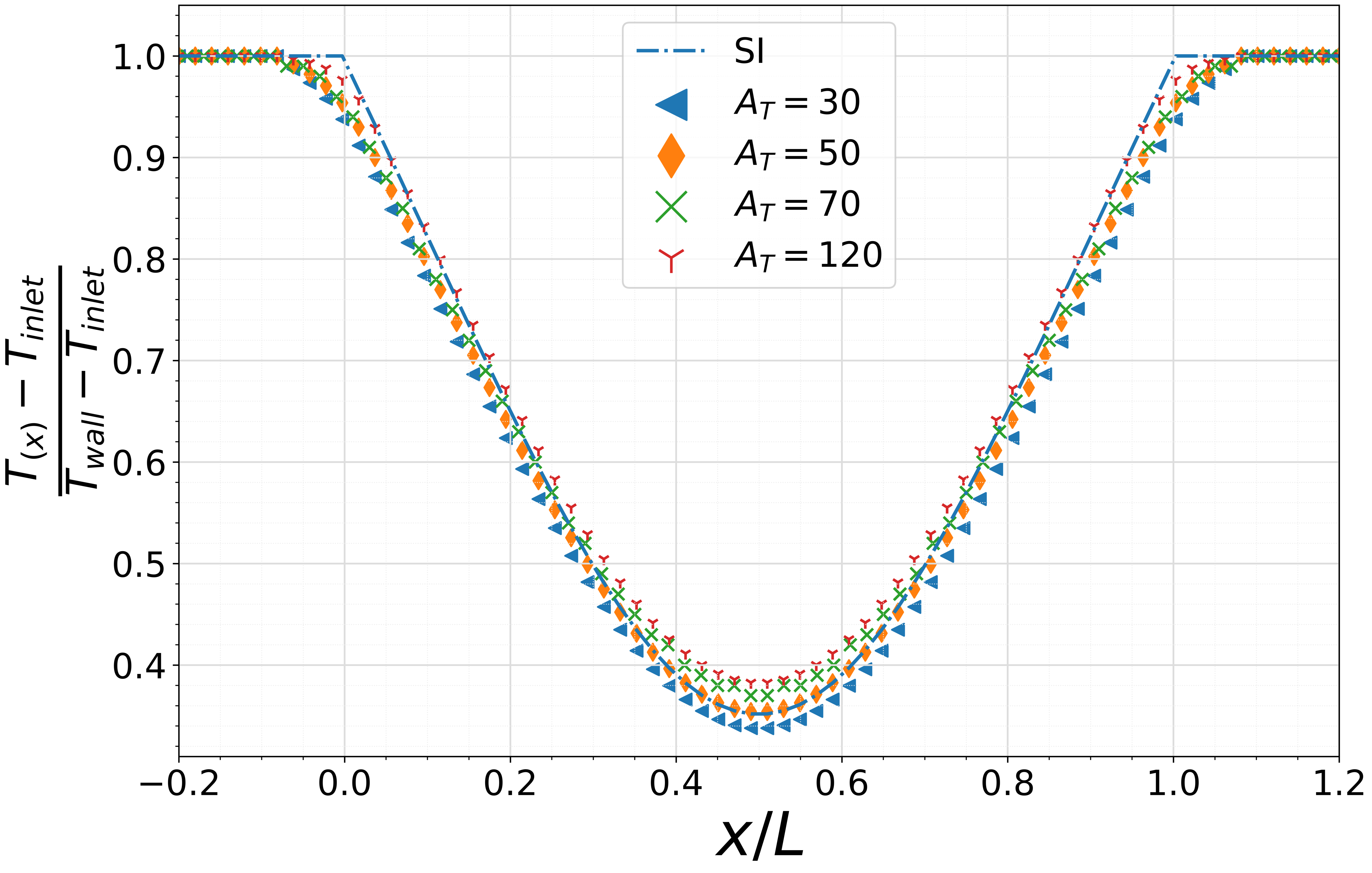}}
    \caption{Temperature profiles at a cross section of $y=0.9H$ obtained within the DI-method for different choices of the thermal coupling parameter $A_u$ as indicated in curve legends. The Reynolds number is $\Rey=100$. The fluid temperature at the inlet is $\Tinlet=300$ K and the temperature of the walls is set to $\Twall=400$ K. For $A_T=70$, a good agreement between the diffuse interface simulation and the sharp interface solution (dashed line) is observed.}
	\label{fig:DI-vs-SI-diff_AT}
\end{figure}

\begin{figure}[t]
	\centering
  \hideimage{  \includegraphics[width=0.9\linewidth]{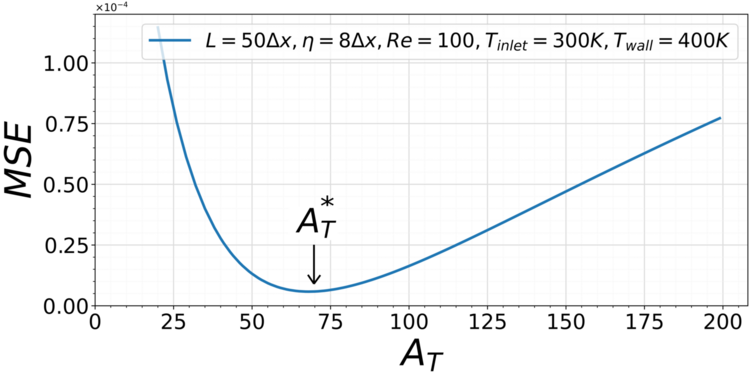}}
   \caption {Mean standard error (MSE) versus coupling parameter $A_T$. Each data point corresponds to a separate simulation using exactly the same channel geometry and the same inlet velocity profile. Only $A_T$ is varied from simulation to simulation. Both the walls are kept at the same constant temperature of $\Twall=400$ K. The fluid enters the channel at a temperature of $\Tinlet = 300$ K. All other simulation parameters are kept constant. As seen from the plot, MSE exhibits a clear minimum upon variation of $A_T$. This optimum value is referred to with $\ATstar$.}
	\label{fig:A_T-vs-MSE}
\end{figure}

To investigate the effect of the coupling parameter $A_T$ on heat transport, we have performed simulations for different values of $A_T$ while keeping all other parameters such as $L,\, H$ and $\eta$ constant. As mentioned earlier, a fully developed parabolic velocity profile has been imposed at the inlet. Figure~\ref{fig:DI-vs-SI-diff_AT} compares the thus obtained results for temperature field with the sharp interface solution. This plot suggests that it is possible to increase the accuracy of the result by a variation of $A_T$. For a quantitative analysis, we compute, for each simulated temperature profile, the mean standard error in temperature, MSE, via Eq.~(\ref{eq:mean-standard-error-f}), where we identify $f(x)\equiv T(x,y)$ for the selected cross section, $y$. Repeating simulations for different values of $A_T$, we thus obtain the dependence of MSE on the coupling parameter $A_T$. Results obtained via this procedure are depicted in Fig.~\ref{fig:A_T-vs-MSE}. Again, and in qualitative similarity to Fig.~\ref{fig:A_u-vs-MSE}, the plot shows the existence of an optimal coupling constant, where the deviation between the diffuse interface simulations and the SI-solution is minimized. As in the case of $A_u$, we use this minimum error criterion to define an optimal coupling parameter, $\ATstar$.

To investigate the effect of the coupling parameter $A_T$ on heat transport, we have performed simulations for different values of $A_T$ while keeping all other parameters such as $L,\, H$ and $\eta$ constant. As mentioned earlier, a fully developed parabolic velocity profile has been imposed at the inlet. Figure~\ref{fig:DI-vs-SI-diff_AT} compares the thus obtained results for temperature field with the sharp interface solution. This plot suggests that it is possible to increase the accuracy of the result by a variation of $A_T$. For a quantitative analysis, we compute, for each simulated temperature profile, the mean standard error in temperature, MSE, via Eq.~(\ref{eq:mean-standard-error-f}), where we identify $f(x)\equiv T(x,y)$ for the selected cross section, $y$. Repeating simulations for different values of $A_T$, we thus obtain the dependence of MSE on the coupling parameter $A_T$. Results obtained via this procedure are depicted in Fig.~\ref{fig:A_T-vs-MSE}. Again, and in qualitative similarity to Fig.~\ref{fig:A_u-vs-MSE}, the plot shows the existence of an optimal coupling constant, where the deviation between the diffuse interface simulations and the SI-solution is minimized. As in the case of $A_u$, we use this minimum error criterion to define an optimal coupling parameter, $\ATstar$.

\begin{figure}[t]
	\centering
\hideimage{\includegraphics[width=1.0\linewidth]{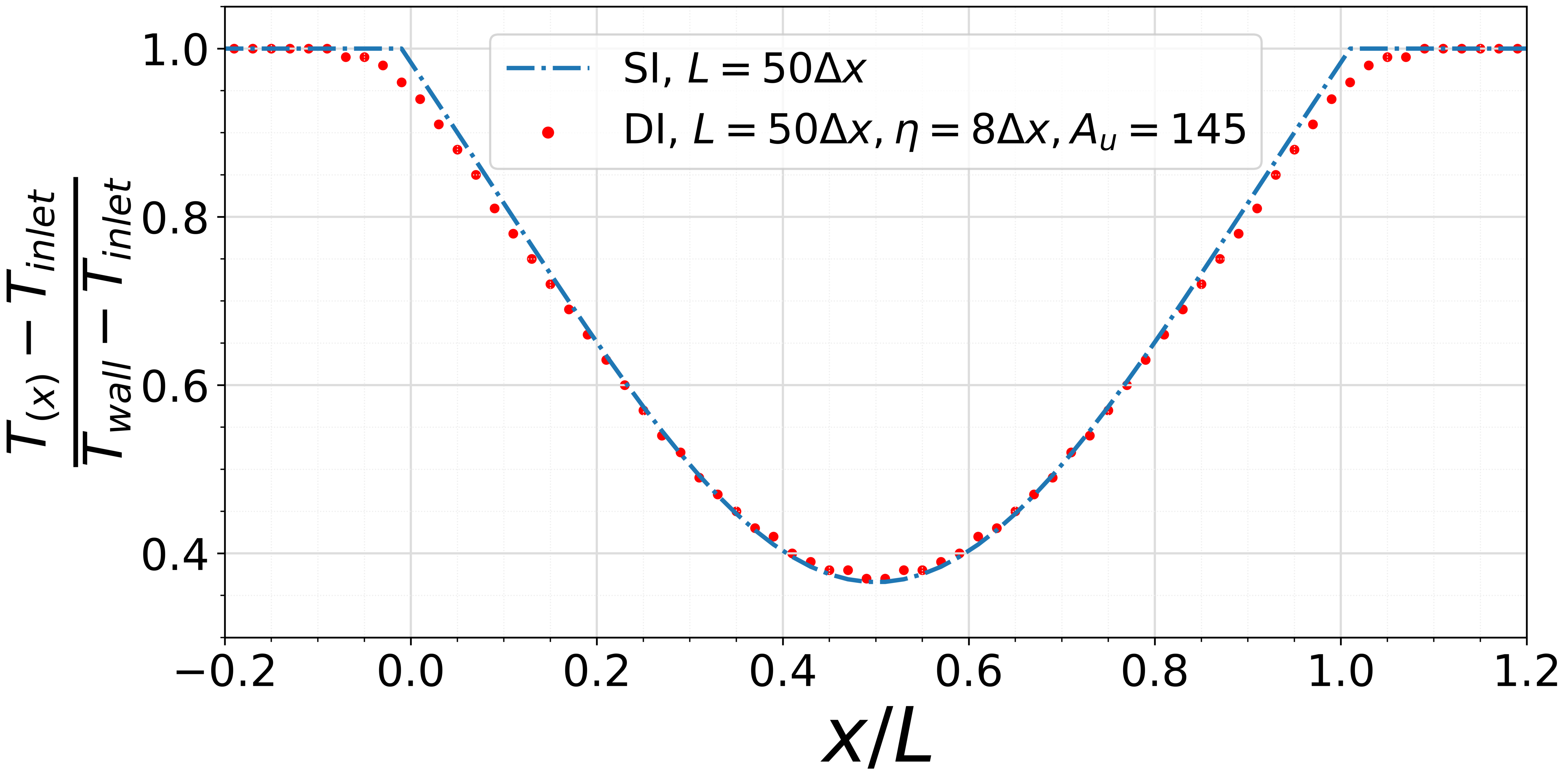}}
    \caption {Simulation results for temperature profile obtained within the diffuse-interface method and the sharp interface solution. The data correspond to a cross section at a distance $y=0.9H$ from the inlet.}
	\label{fig:DI-vs-SI-T}
\end{figure}

To examine the quality of the data, Fig.~\ref{fig:DI-vs-SI-T} compares the temperature profile obtained within the phase-field based approach for $\ATstar$ to the sharp interface solution. It can be confidently stated that a very good agreement exists between the two profiles within the bulk region of the fluid. 

\begin{figure}[t]
	\centering
	\hideimage{\includegraphics [width=1.0\linewidth]{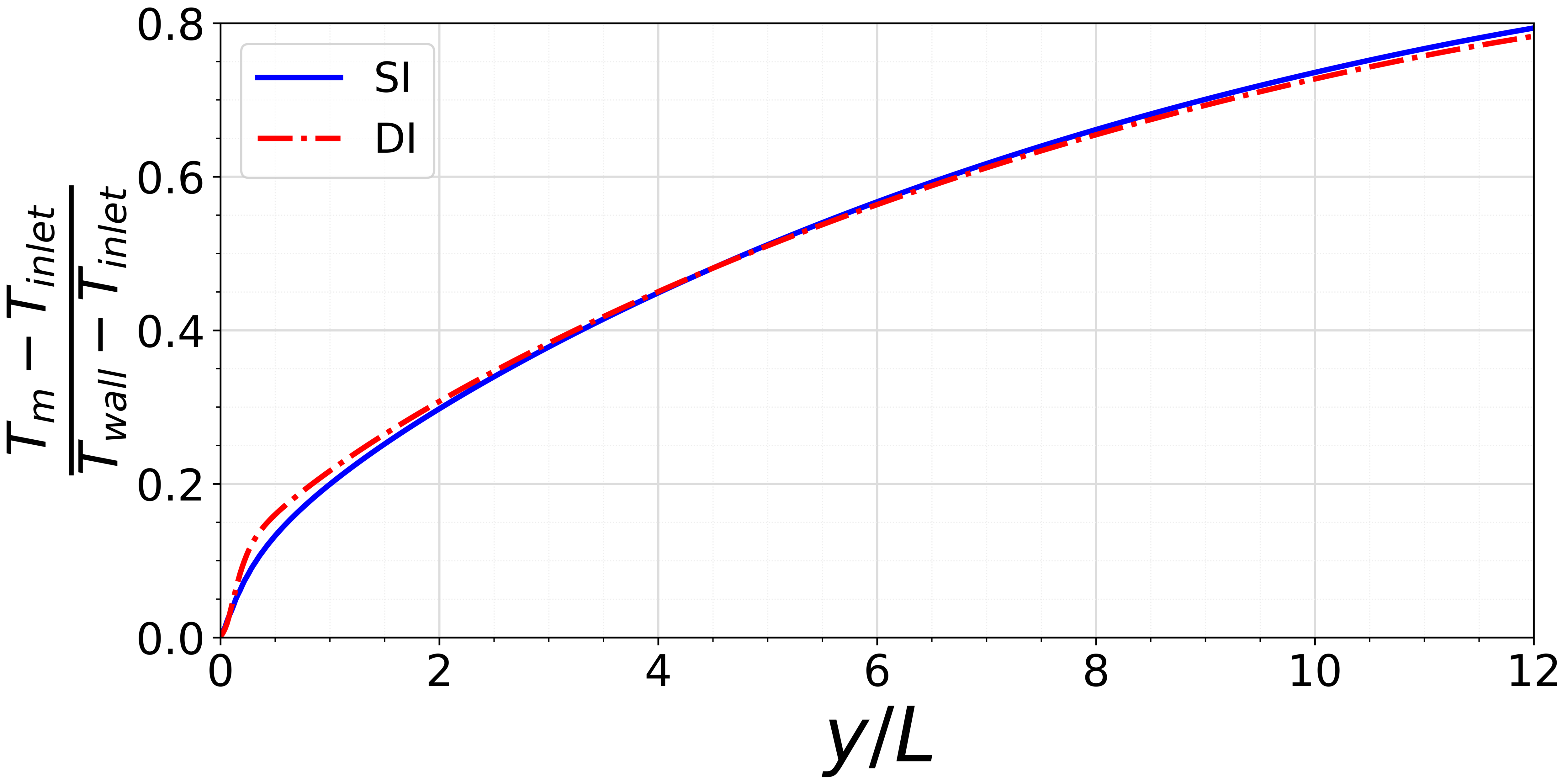}}
    \caption {A comparison of the mean cross sectional temperature (evaluated via Eq.~(\ref{eq:Tmean-def} and normalized) obtained within the diffuse interface (DI, dashed-dotted line) and the sharp interface (SI, solid line) methods. The value of the DI-coupling parameter is $A_T=70$. The Reynolds number is $\Rey=100$. The wall temperature is $\Twall=400$ K and inlet temperature of the fluid is $\Tinlet=300$ K. The length of the channel is 12 times the wall-to-wall distance, $H=12 L$. There is a good overall agreement between the simulated data and the reference SI-solution. Some deviations, do, however, occur close to the inlet (see the text for a discussion of this issue).}
	\label{fig:SI-vs-DI_Tm}
\end{figure}

As a further test, Fig.~\ref{fig:SI-vs-DI_Tm} compares results for the mean cross sectional temperature, $\Tm(y)$, obtained within the diffuse interface method at the optimal coupling constant $\ATstar$ with the corresponding sharp interface reference data. The Reynolds number is equal to $\Rey=100$. The wall temperature is maintained at a constant value of $\Twall=400$ K, while the inlet temperature is $\Tinlet=300$ K. The distance between the walls is $L=50 \Delta x$ and the thickness of the diffuse interface is $\eta=8\Delta x$. It is also visible from the data that deviations between the DI and SI-results for mean temperature are larger closer to the inlet (smaller values of $y$). This can be attributed to the fact that, at the inlet, a cold fluid comes into contact with a hot wall, which gives rise to a large temperature gradient. Generally, dealing with large gradients poses a challenge and enhances the deviation between alternative numerical schemes such as the present DI and SI methods.

\begin{figure}[t]
	\centering
	\hideimage{\includegraphics [width=1.0\linewidth]{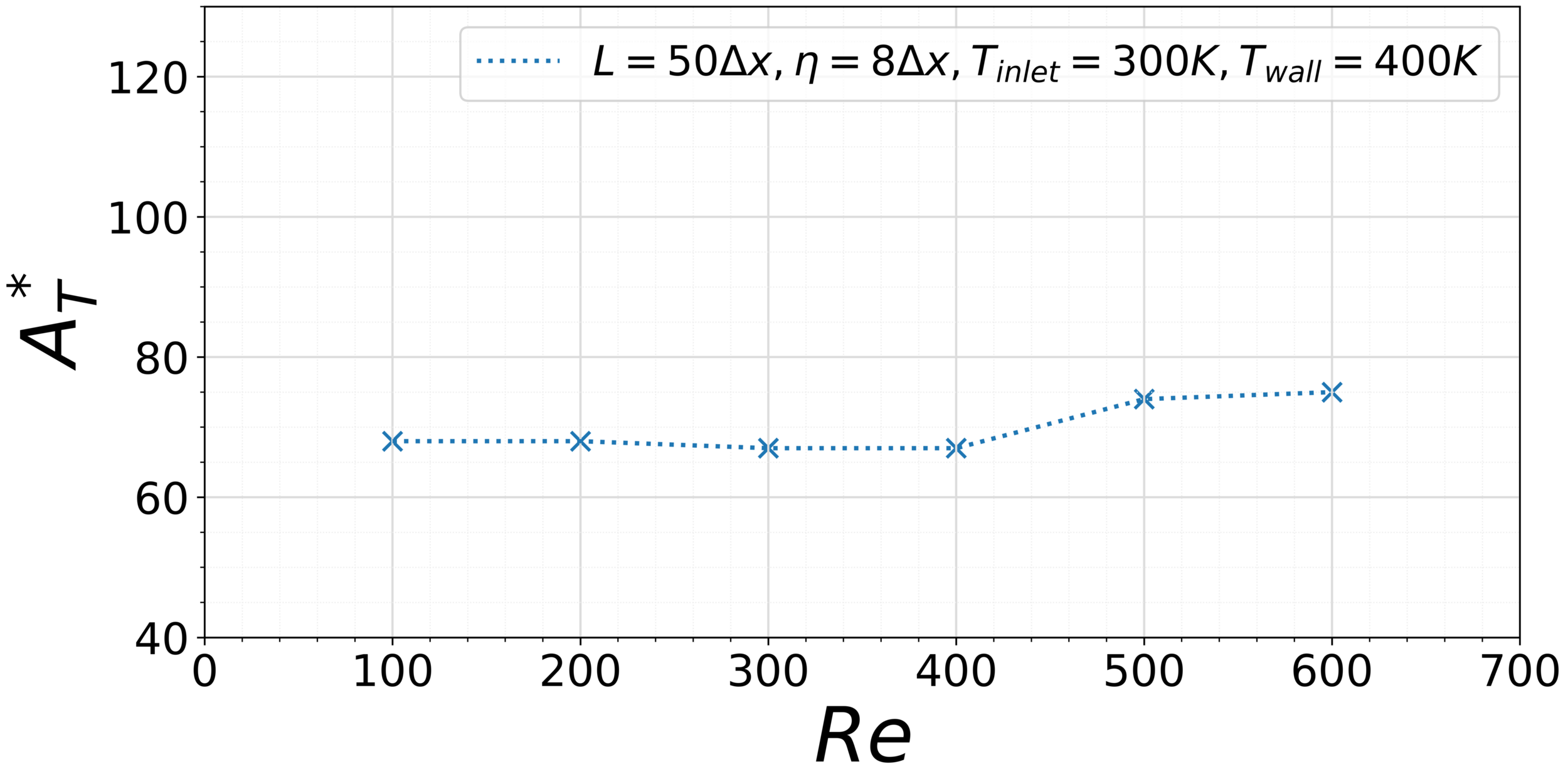}}
    \caption{The coupling parameter for heat transfer $\ATstar$ versus Reynolds number. All other simulation parameters including wall temperature, inlet temperature, $L$, $H$ and $\eta$ are kept constant in all the simulations performed for this purpose.}
	\label{fig:A_T-vs-Re}
\end{figure}

\begin{figure}[t]
	\centering
\hideimage{\includegraphics [width=1.0\linewidth]{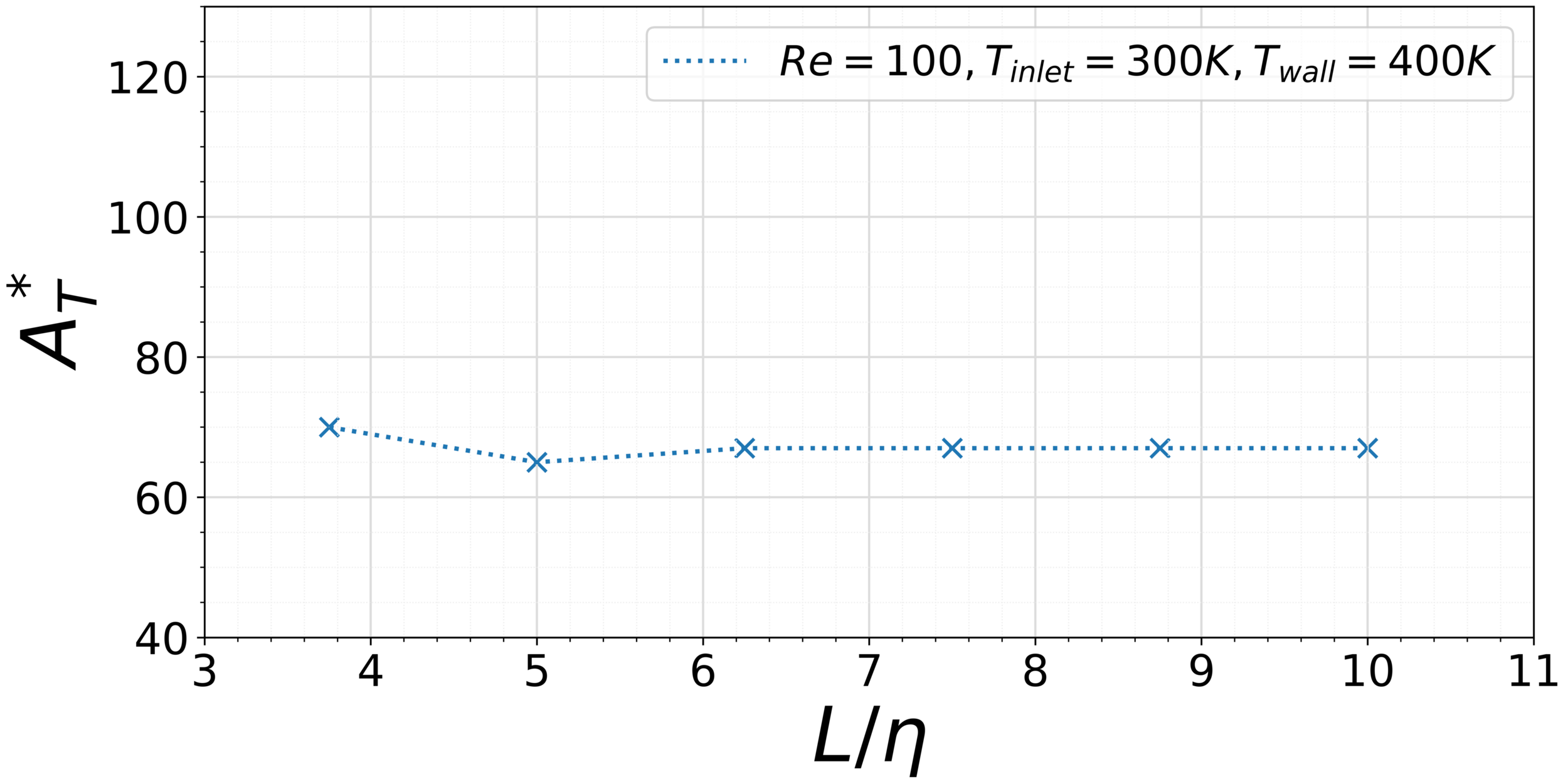}}
    \caption {The optimal coupling parameter for heat exchange $\ATstar$ versus the non-dimensional ratio of the channel width to interface thickness, $L/\eta$ at a constant Reynolds number of $\Rey=100$. Similar to the behavior of $\ATstar$ (Fig.~\ref{fig:Au*-vs-L/Eta}), $\ATstar$ first decreases with $L/\eta$ but then reaches a size-independent plateau for $L \geq 5 \eta$.}
	\label{fig:A_T-vs-L/Eta}
\end{figure}

Next we examine the possible influence of the flow velocity on the optimal coupling parameter for heat transfer, $\ATstar$. For this purpose, we use the same simulation setup as described above but vary systematically the inlet flow velocity such that the Reynolds number changes in the range $\Rey=100-600$. Again, as in the case of Fig.~\ref{fig:A_u-vs-MSE}, a two dimensional parameter space is sampled for this investigation. Results of these simulations are shown in Fig.\ref{fig:A_T-vs-Re} and indicate that, similar to the case of $\Austar$, the optimal coupling parameter for heat exchange is also insensitive to variations of the flow velocity. A similar study, now focusing on the effect of system size, further reveals that $\ATstar$ is robust with regard to variations of $L/\eta$, see Fig.\ref{fig:A_T-vs-L/Eta}.

\begin{figure}[t]
	\centering
\hideimage{\includegraphics [width=1.0\linewidth]{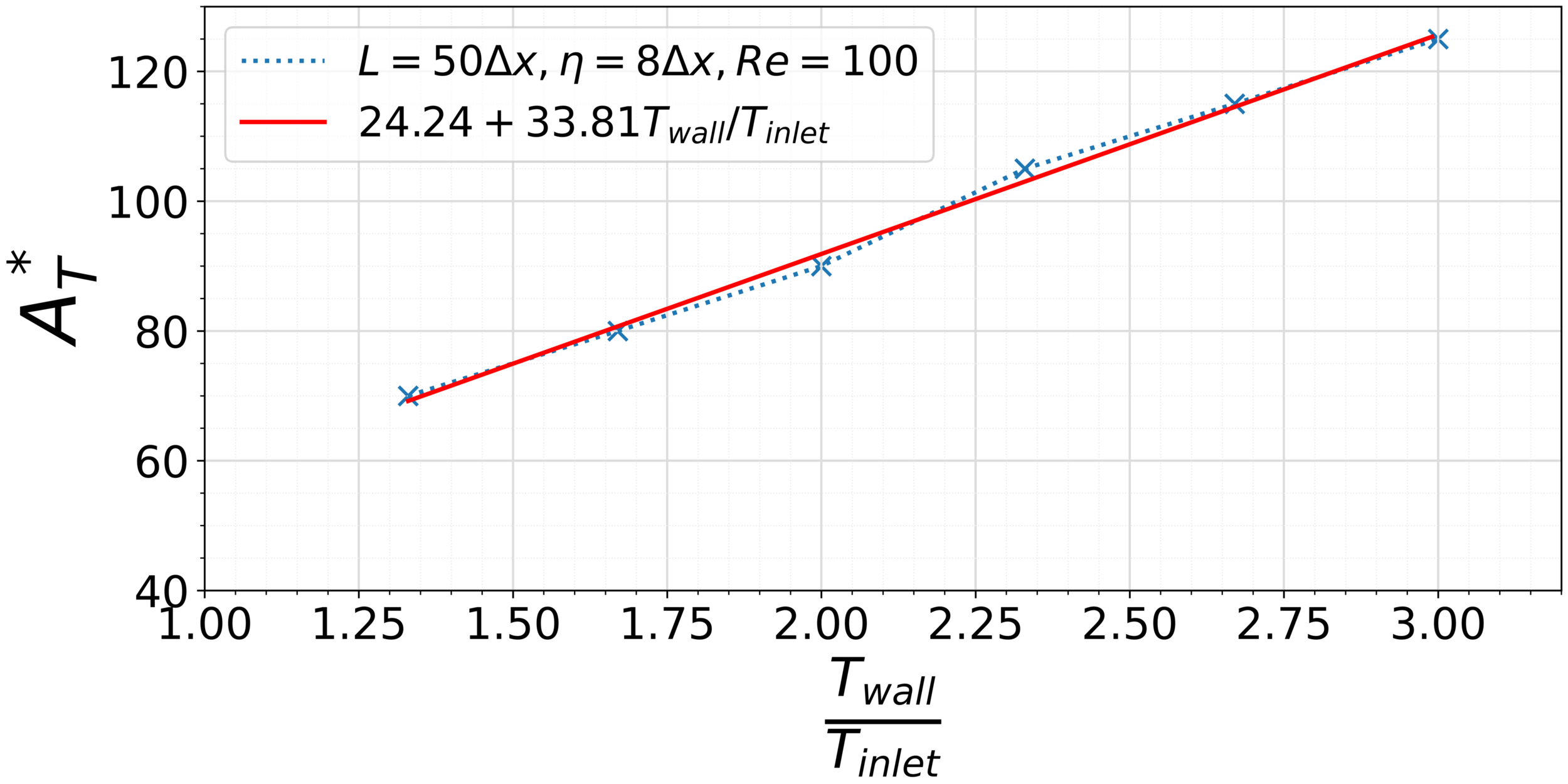}}
    \caption {The effect of the fluid-solid temperature difference on the optimal coupling parameter, $\ATstar$. The plot shows $\ATstar$ versus the temperature ratio, $\Twall/\Tinlet$, for diffuse interface simulations. All other parameters such as $\Rey$, $L$, $H$ and $\eta$ are kept constant. The wall temperature is incrementally raised from $\Twall=400$ K to $900$ K and, for each selected temperature, the optimal coupling parameter for heat exchange is determined by sampling the $A_T$-space. As seen from the plot, $\ATstar$ can be approximately described by a linear function. A least square fit yields, $\ATstar=24.24+33.81{\Twall}/{\Tinlet}$. This relation can be used to estimate the value of $\ATstar$ at any other wall temperature in the investigated interval.}
	\label{fig:A_T-vs-Tw}
\end{figure}

As the next question, we address the possible influence of the wall-fluid temperature difference on $\ATstar$. This is a particularly important question, since the main purpose of $\ATstar$ is to tune the heat exchange between the fluid and the wall. For this purpose, a systematic investigation is conducted, while sampling a two-dimensional parameter space spanned by $A_T$ and $\Twall$. The variation of the fluid-solid temperature difference is achieved by keeping the inlet fluid temperature at $\Tinlet=300$ K and varying the wall temperature in the range of $\Twall\in\{400, 900\}$ K. Results of these simulations are shown in Fig.~\ref{fig:A_T-vs-Tw}. It turns out that, unlike the negligible effect of flow velocity and channel width (at sufficiently large $L/\eta$), wall temperature has significant effect on the optimal coupling parameter, $\ATstar$. Interestingly, as shown in Fig.~\ref{fig:A_T-vs-Tw}, a linear fit describes quite well the dependence of $\ATstar$ on the wall temperature. The thus obtained fit function can be used to estimate the value of $\ATstar$ in the investigated range of $\Twall=400-900$ K.

\subsection{\blue{Convergence rate study}}

\blue{To evaluate the accuracy and convergence characteristics of the proposed method, we consider a classical benchmark problem with an analytical solution: steady, incompressible, thermally, and hydrodynamically fully developed Poiseuille flow with heat transfer between two vertical parallel plates. The plates, located at $x = 0$ and $x = L$, are maintained at a constant wall temperature $T_w$, while a Newtonian fluid (air) enters the channel at a uniform inlet temperature $T_\text{in}$. The channel has a length of $H = 30L$, which ensures that fully developed flow conditions are achieved. The Reynolds number is set to $\text{Re} = 100$, based on the mean inlet velocity $\bar{u}_\text{in}$, the characteristic length $L$, and the kinematic viscosity $\nu$.}
\blue{The analytical solution for the velocity profile is parabolic,
\begin{equation}
    \blue{u(x) = 6 U_m \frac{x}{L} \left( 1 - \frac{x}{L} \right).}
\end{equation}
where $U_m$ is the mean velocity at the inlet. A nondimensional temperature variable $\hat{T}$ is introduced above, subject to the following boundary conditions,
\begin{align}
    \hat{T}(x, 0) &= 1,\\
    \hat{T}(0, y) &= \hat{T}(L, y) = 0,
\end{align}
Neglecting axial conduction—valid for sufficiently high Péclet numbers, $\mathrm{Pe} = \frac{U_m L}{\alpha}$—and assuming constant thermophysical properties, the analytical solution for the temperature field can be expressed as a Fourier sine series,
\begin{equation}
    \blue{\hat{T}(x, y) = \sum_{n=1}^{\infty} \frac{4}{n\pi} \sin\left( \frac{n\pi x}{L} \right) }
    \blue{\exp\left[ -\left( \frac{n\pi}{L} \right)^2 \frac{y U_m}{\alpha} \right],}
    \label{eq:graetz_solution}
\end{equation}
where $\alpha$ is the thermal diffusivity. This solution satisfies the imposed thermal boundary conditions and describes the evolution of the temperature profile as a function of both the wall-normal coordinate $x$ and the axial coordinate $y$. It captures the thermal development in the entrance region of the channel and serves as a reference for validating numerical temperature fields in similar flow configurations.}

\begin{figure}[t]
    \centering
    \includegraphics[width=0.8\linewidth]{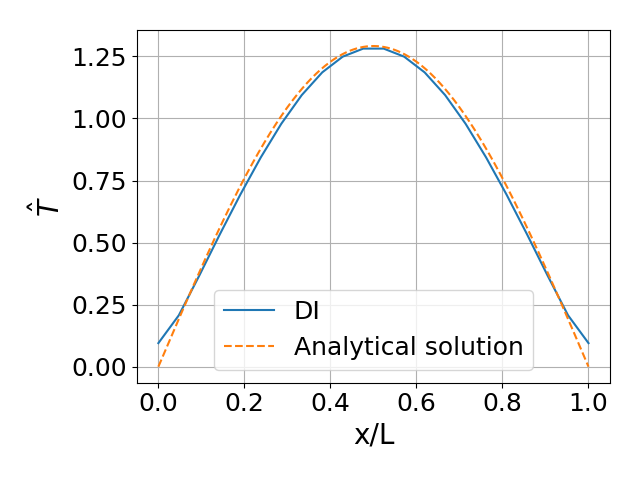}
    \caption{\blue{Comparison of non-dimensional temperature profiles obtained from the DI-based simulation and analytical solution at $x = 0.99H$.}}
    \label{fig:T-DI-Analytic}
\end{figure}

\blue{Figure~\ref{fig:T-DI-Analytic} compares the numerical and analytical profiles of temperature. This profile is extracted at $x = 0.99H$, where the flow is fully developed.}

\begin{figure}[t]
    \centering
    (a) \includegraphics[width=0.8\linewidth]{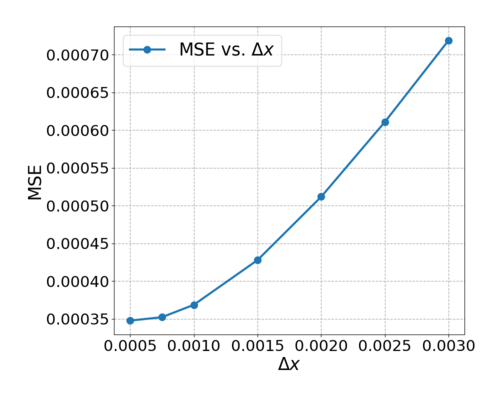} \\
    (b) \includegraphics[width=0.8\linewidth]{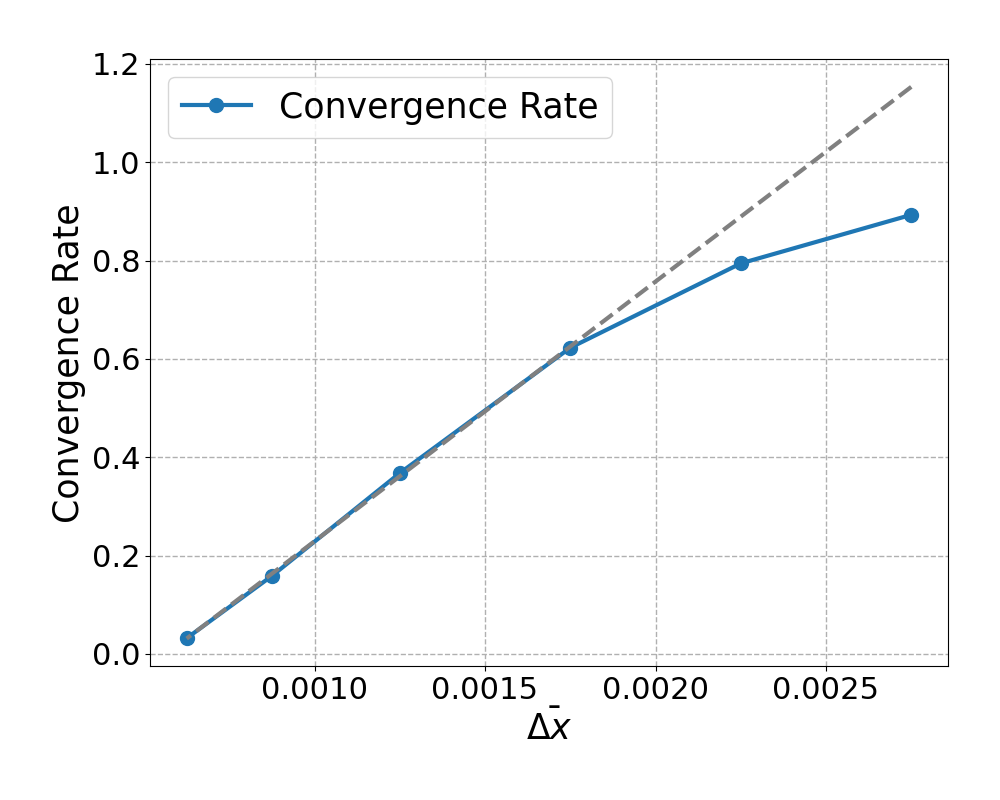}
    \caption {\blue{(a) Mean Square Error (MSE) versus grid spacing $\Delta x$. MSE is evaluated from the difference between simulated data and analytical solution. Starting from large $\Delta x$, the error decreases rapidly as the grid size is reduced. Beyond a certain resolution, the error approaches a plateau. This plateau arises from model-dependent systematic errors such as expansion due to tiny changes in temperature, whereas the analytic solution assumes a constant density. Despite this, the overall accuracy is quite high, with the relative numerical MSE being on the order of $10^{-4}$. (b) Convergence rate versus $\Delta x$ as obtained from the data shown in panel (a), using Cov.~Rate=$\frac{\log(MSE_2/MSE_1)}{\log(\Delta x_2/\Delta x_1)}$. The decrease of the convergence rate reflects the above-mentioned fact (see panel (a)) that the numerical accuracy has reached a high level, so that systematic errors become visible.}}
\label{fig:convergence_rate}
\end{figure}

\blue{To assess the convergence rate, simulations are conducted using various grid resolutions. The mean square error between the numerical and analytical solutions is calculated for the temperature profile at the cross-section located at $x = 0.99H$. Figure~\ref{fig:convergence_rate} presents how the MSE varies with grid spacing $\Delta x$. At large grid spacings, the error decreases rapidly as the grid is refined. Beyond a certain resolution threshold, however, it converges to a plateau. This saturation arises from model-specific systematic deviations—such as thermal expansion caused by minute temperature fluctuations—while the analytical solution presumes a constant density. Despite this limitation, the numerical accuracy remains high, with a relative mean squared error on the order of $10^{-4}$.}

\subsection{Reactive flow}
In this section, we present the results of simulations for a reacting gas flow in a number of geometries including the flow through a particle bed. For this purpose, our home-made open-source simulation tool OpenPhase Academic (www.openphase.rub.de) is coupled to the open-source chemical kinetics software CANTERA (https://cantera.org), which provides the thermodynamic and transport properties of the reacting chemical species and information on the proper reaction mechanisms. To demonstrate the capability of the code in accounting accurately for the relevant chemical reactions, a number of simulations are performed and the outcomes are validated both against experiments and versus the known literature data. The effective transport properties of the mixture such as viscosity and thermal conductivity are computed according to the mixture model introduced in~\cite{kee2005},
\begin{align}
	\mu &=  \sum_{k=1}^{N} \frac{\mu_k X_k}{\sum_{j=1}^{N} \Phi_{kj}X_j},  \label{eq:muT_mix} \\
	\Phi_{kj}& =  \frac{\Big(1+\sqrt{ (\frac{\mu_k}{\mu_j} \sqrt{\frac{M_j}{M_k}})}\Big)^2}{\sqrt{8}\sqrt{1+M_k/M_j}}, \label{eq:phi_mix} \\
	\lambda & =  0.5\Big(\sum_{k=1}^{N} X_k \lambda_k + \frac{1}{\sum_{k=1}^{N}X_k/\lambda_k } \Big),  \label{eq:conduct_mix} 
\end{align}
where $\mu_k$ and $\lambda_k$ denote the dynamic shear viscosity and the thermal conductivity of the species $k$, respectively.

In all the reactive gas simulations reported below, a mixture of methane and air is utilized. For details of the reaction, we use the standard \enquote{BFER} mechanism, which is recognized for its applicability in studying laminar premixed flames. It involves six species (CH4, O2, N2, CO, CO2, and H2O) and comprises two reaction steps~\cite{franzelli2012}. Using this reactive system, the proposed methodology is validated against available literature data for a 1-D flame (sec.~\ref{sec:1D-flame}) and flame-wall interactions in a 2-D setup (sec.~\ref{sec:2D-flame}). In sec.~\ref{sec:flame-in-packed-bed}, a comparison to experiments in the case of a flame in a bed of inert particles is presented. After these validations, the ,methodology is applied to study the dynamics of flame front in an assembly of complex-shaped particles (\ref{sec:flame-in-comlex-packed-bed}).


\subsubsection{Validation via a 1-D laminar flame}
\label{sec:1D-flame}

Flame propagation is a complex phenomenon involving physical and chemical processes such as thermal interactions between the flame and the walls, chemical reaction among species within the flame, heat exchange and momentum transfer between the flow field and the flame~\cite{kim2006}. In this context, the study of relatively simple 1-D flames is an important validation step as it provides key characteristics of premixed mixtures, including flame thickness, propagation speed, mass and heat diffusion, and adiabatic flame temperature. Here, we use a mixture of methane-air at an equivalence ratio of $\phi=1$. The inlet temperature of the methane-air mixture is chosen to be 300 K. This simulation of the 1-D flame is performed using a fine grid in order to capture well the spatial structure of the flame and the details of its sharp front. 

To establish a stationary flame in the simulation domain, the mass flow rate of the methane-air mixture entering the channel is adjusted at every time step as to compensate the backward motion of the flame towards the inlet. For this purpose, the consumption rate of the fuel (here methane) at the flame front is estimated. This information is then used to evaluate the flow velocity at the inlet in order to replenish the consumed fuel. This approach proved to be very effective and converges rapidly towards a stationary flame.

From this study, the propagation speed of the flame front is found to be $S_\text{F}=0.407$ m/s, which is in good agreement with the previous studies. For example, in Refs.~\cite{Hosseini2023} and \cite{kim2006}, the flame speed for the same input parameters has been reported to be $S_\text{F}=0.408$m/s and $0.4$ m/s, respectively.

The thermal flame thickness is estimated using the standard relation~\cite{Hosseini2023, kim2006}, 
\begin{equation}
	\delta_T= \frac{\Tadiab-\Tinlet}{dT/dx\vert_\text{max}},
	\label{eq:thermal-thickness}
\end{equation}
where $\Tadiab$ is the adiabatic flame temperature. For the present simulation of a 1-D flame of a methane-air mixture, the thermal flame thickness is thus estimated to be  $\delta_T=\SI{342}{\micro\metre}$. This value is close to the reported value of $\delta_T=328\mu$m~\cite{Hosseini2023}.

\blue{In view of such a thin flame front, a special attention must be payed to the choice of grid spacing. Therefore, we have conducted a grid-independence study and have found that for $\Delta x < 30~\mu$m results are essentially independent of the grid spacing.} This is in line with the recommendation that the grid size shall be at least ten times smaller than the thermal thickness of the flame front~\cite{kim2006}.

\begin{figure}[t]
	\centering
	\hideimage{\includegraphics [width=0.8\linewidth]{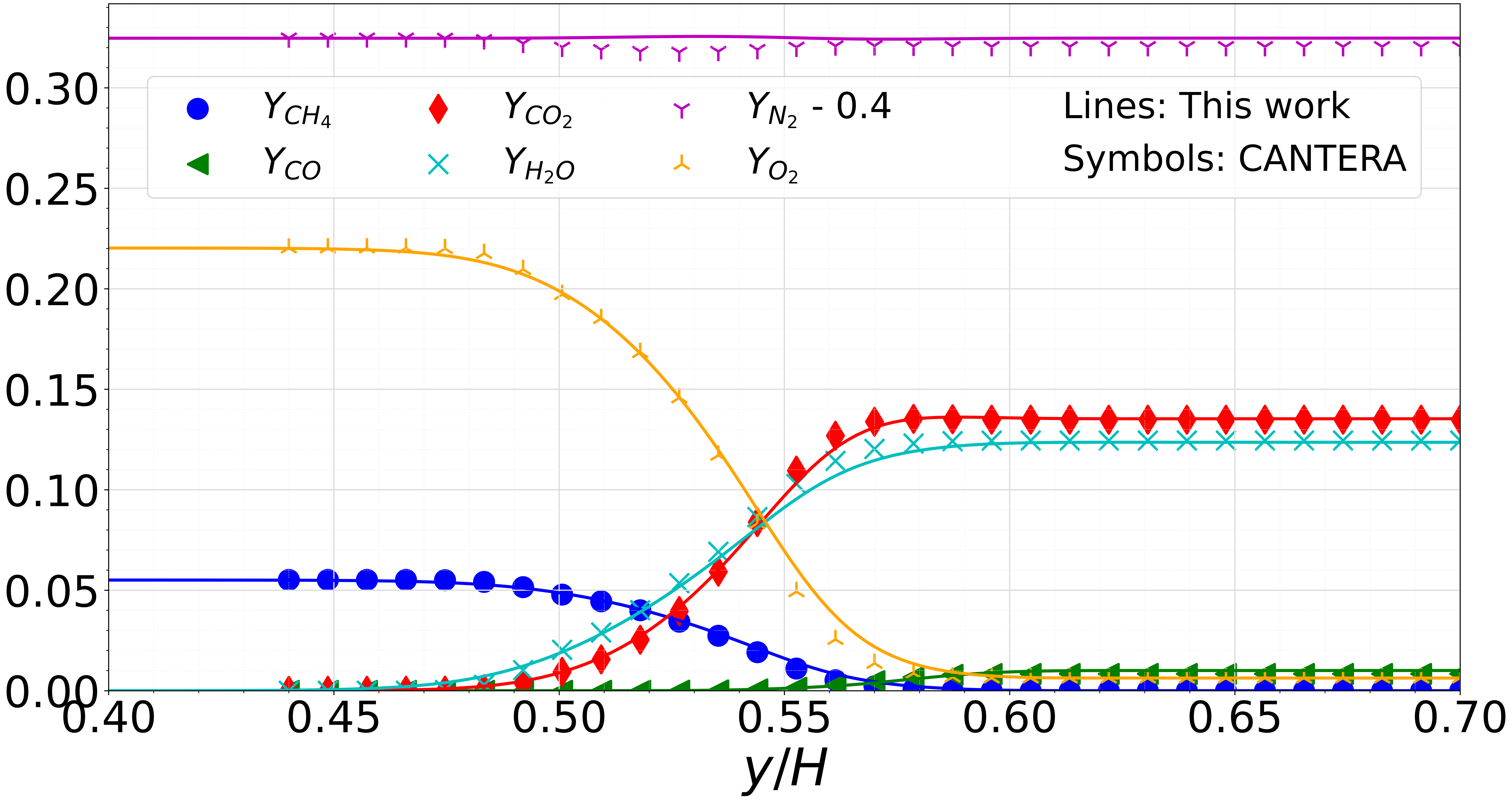}}
    \caption{Mass fractions of the species participating in the 1-D flame of methane-air mixture using BFER reaction mechanism. To put the data for the nitrogen mass fraction in the same plot range, we show Y$_\text{N2}-0.4$. Lines correspond to the results obtained within the proposed hybrid LB-FD-PF approach, while symbols show the data obtained via CANTERA.}
	\label{fig:mass-frac-1d}
\end{figure}

\begin{figure}[t]
	\centering
\hideimage{\includegraphics [width=0.8\linewidth]{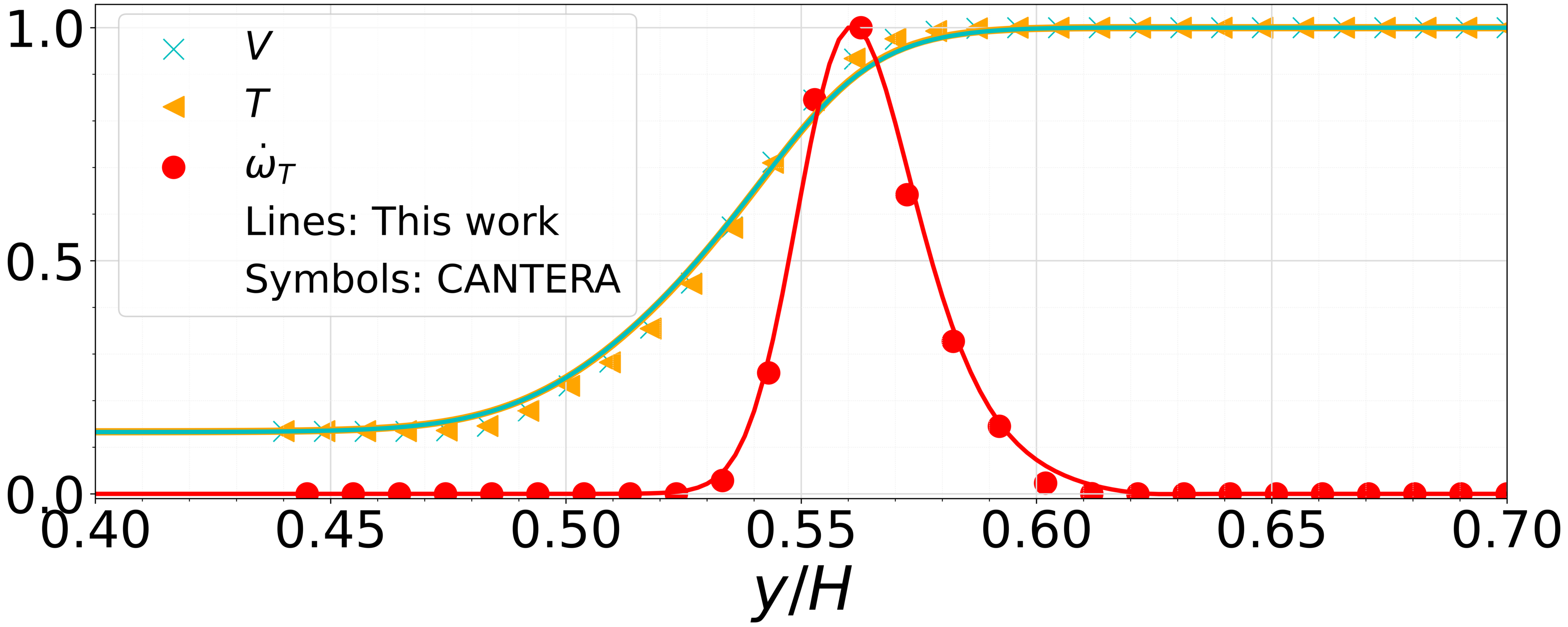}}
    \caption{Temperature, velocity and heat release of the mixture in the 1-D flame simulation along the flow direction. All the values are normalized with the maximum value of each variable. The maximum temperature, velocity and heat release are $T_\text{max}=2252$ K, $\umax=3.07$ m/s and $\dot\omega_{T_\text{max}}=7.32\times 10^9\; \text{watt/m}^3$, respectively. Lines correspond to the results obtained within the proposed hybrid LB-FD-PF approach, while symbols show the data obtained via CANTERA.}
\label{fig:v-T-rho-1d}
\end{figure}

Next we provide a more detailed comparison of various quantities within the 1-D flame. For this purpose, we mention that the chemical kinetics library CANTERA not only provides various kinetic data required for chemical reactions in our numerical solver but also has the capability to independently simulate 1-D flames. Using this feature, we compare in Fig.~\ref{fig:mass-frac-1d} the results obtained within the present hybrid simulation approach with the data provided by CANTERA regarding the concentrations of various species involved in the reaction and their spatial variation across the flame front. Similarly, Fig.~\ref{fig:v-T-rho-1d} compares temperature, velocity and the rate of heat release across the flame front between the two numerical tools.  As seen from these comparisons, the data obtained within the proposed simulation methodology agree well with the reference benchmark calculations.


\subsubsection{2-D wall-bounded flame}
\label{sec:2D-flame}

\begin{figure}[t]
	\centering
\hideimage{\includegraphics[width=1.0\linewidth]{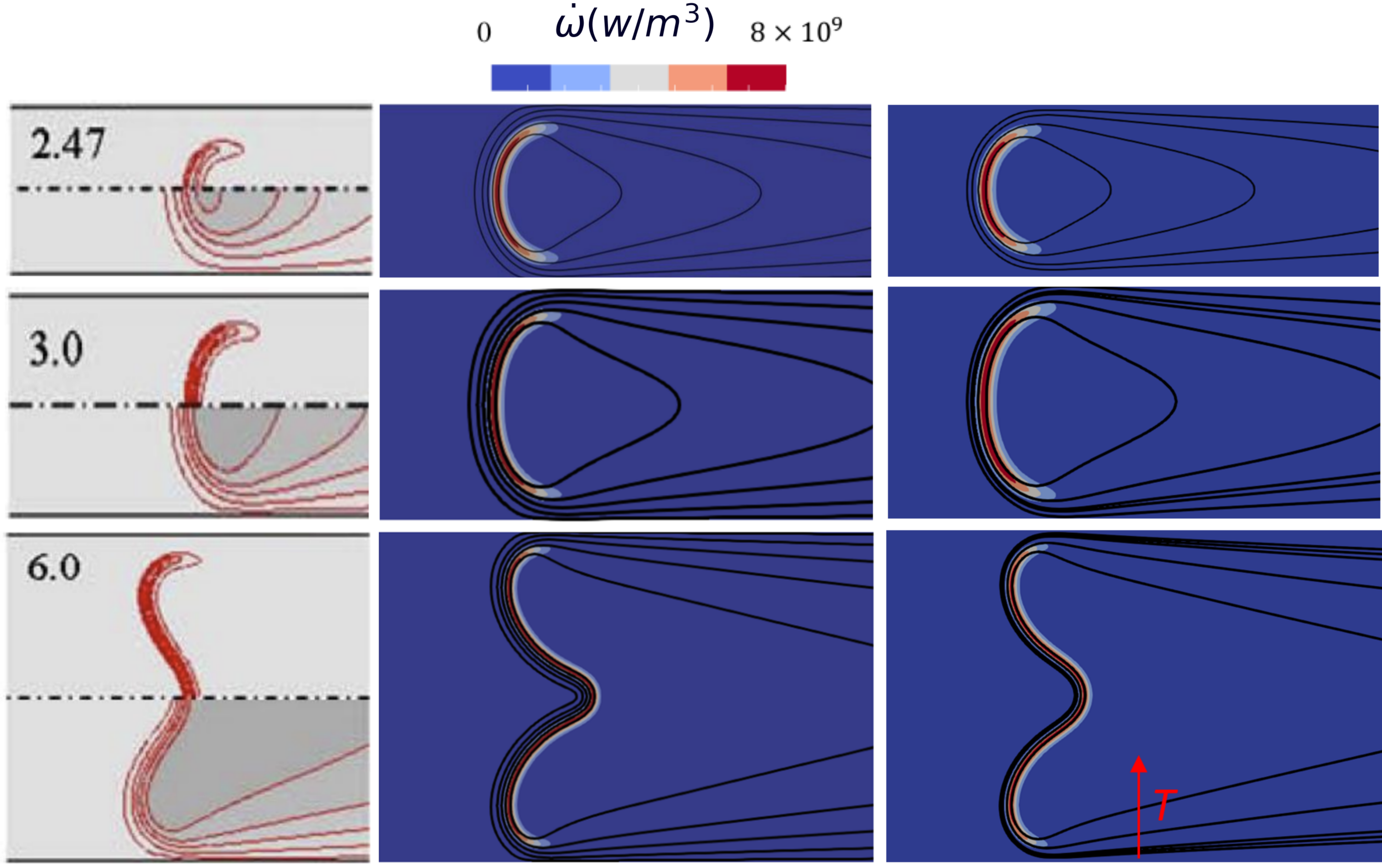}}
    \caption {A comparison of 2-D flame structures obtained within the present work (right column) and two independent studies from the literature, (middle column) reproduced from [Hosseini et al., Physics of Fluids {\bf 35}, 0153814 (2023)], with the permission of AIP Publishing~\cite{Hosseini2023} and (left column) reproduced with permission from Combustion and Flame {\bf 146}, 283 (2006). Coyright 2006 Elsevier~\cite{kim2006}. Each row shows the data for a channel width, given in mm on the top-left corner of the corresponding row. The lines are contours of constant normalized temperature, $\theta=\frac{T-\Tinlet}{\Tadiab-\Tinlet}\in\{0.1,0.3,0.5,0.7,0.9\}$. The red arrow indicates the direction of temperature raise. The color field stands for the heat generation rate, thus highlighting the position and shape of the flame. In each row, the number on top left corner gives the simulated tube diameter.}
	\label{fig:2D-flame-isowalls-all}
\end{figure}

As the next validation step, we use the same configuration as in subsections \ref{sec:Test-of-isothermal-flow} and \ref{sec:Non-isothermal-flow} to simulate a 2-D flame in a straight channel.
The obtained data on temperature distribution and flame structure are then compared with the simulation results reported in \cite{Hosseini2023} and \cite{kim2006}. To investigate the influence of isothermal walls on the flame structure, three different channel widths of 2.47~mm, 3~mm and 6~mm are selected. Throughout all these simulations, the temperature of the walls is maintained at 300~K. The velocity profile at the inlet is parabolic along the transverse direction, $x$, but independent of the longitudinal coordinate, $y$. To establish a flame at the position $y=H/4$, the first quarter of the domain is filled with a fresh methane-air mixture at $T=300$K (equivalence ratio of $\phi=1$) and the remaining portion of the domain is filled with post-combustion products at $T=2200$K. At the beginning of simulations, pressure waves occur which can disrupt the stabilization process~\cite{schlaffer2013}. This problem becomes more pronounced in the presence of a flame. As already mentioned in section~\ref{sec:Test-of-isothermal-flow}, the use of a non-reflecting pressure outlet boundary condition~\cite{poinsot1992} proved quite useful in counteracting these instabilities. Moreover, the standard no-slip boundary condition is applied at the walls. The time step and grid size are $\num{2.5e-7}$ s and $\num{2.5e-5}$ m, respectively. Figure~\ref{fig:2D-flame-isowalls-all} shows the heat release rate and temperature fields obtained from our simulations, along with two other studies referenced as~\cite{Hosseini2023, kim2006}. In a 2-D flame with isothermal walls, due to the significant heat loss near the walls, the flame begins to extinguish at a certain distance from the walls. As can be inferred from Fig.~\ref{fig:2D-flame-isowalls-all}, this feature is well reproduced in our simulations. 
Moreover, the structure of the simulated flame agrees well with the results of previous works~\cite{Hosseini2023, kim2006}.

\begin{figure}[t]
	\centering
	(a)\\ \hideimage{\includegraphics[width=0.9\linewidth]{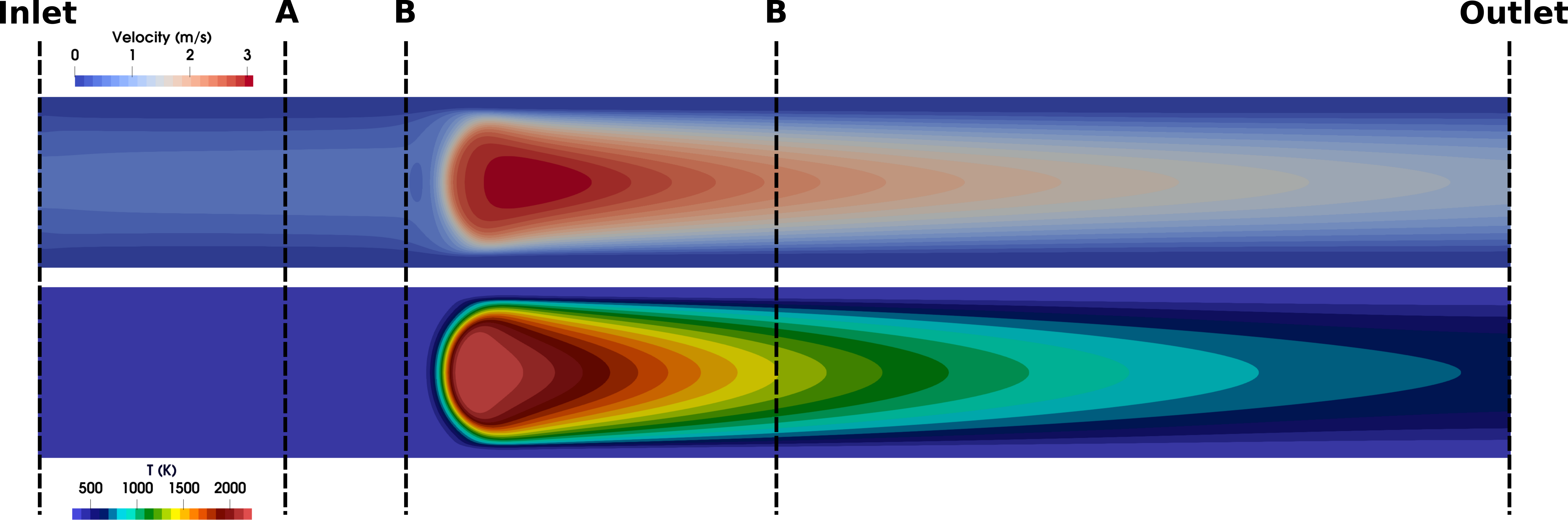}}
	(b)\\\hideimage{\includegraphics[width=0.9\linewidth]{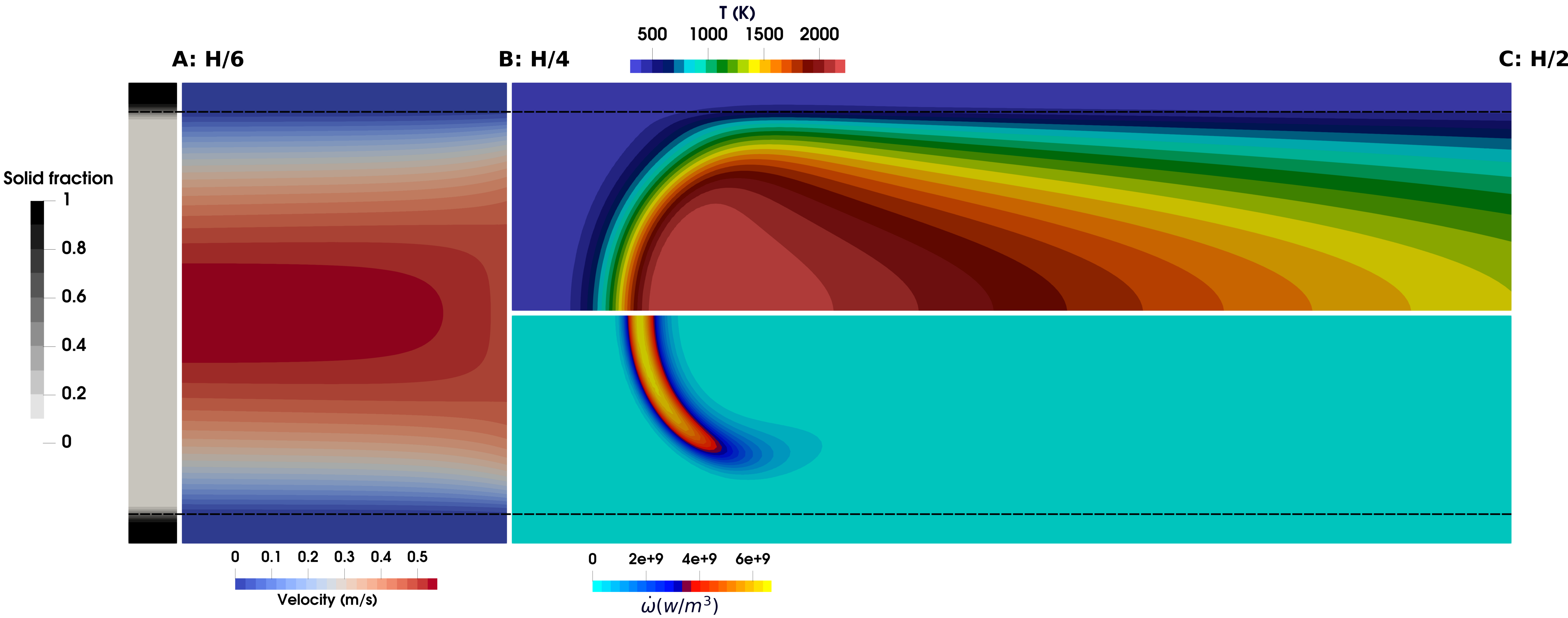}}
    \caption{Top: The velocity and temperature distributions within a 2-D simulation of a flame confined between two parallel walls. The temperature is kept constant at $300$ K for both the walls and for the fluid at the inlet. Bottom: A closer view of the flame front, where substantial gradients in both velocity and temperature fields occur. The dashed line highlights the equivalent sharp interface position of the wall, defined here via $\varphi(x_\text{wall})=0.5$.}
	\label{fig:T-V-Flame-2D-DI}
\end{figure}

\begin{figure}[t]
	\centering
    \hideimage{\includegraphics[width=0.7\linewidth]{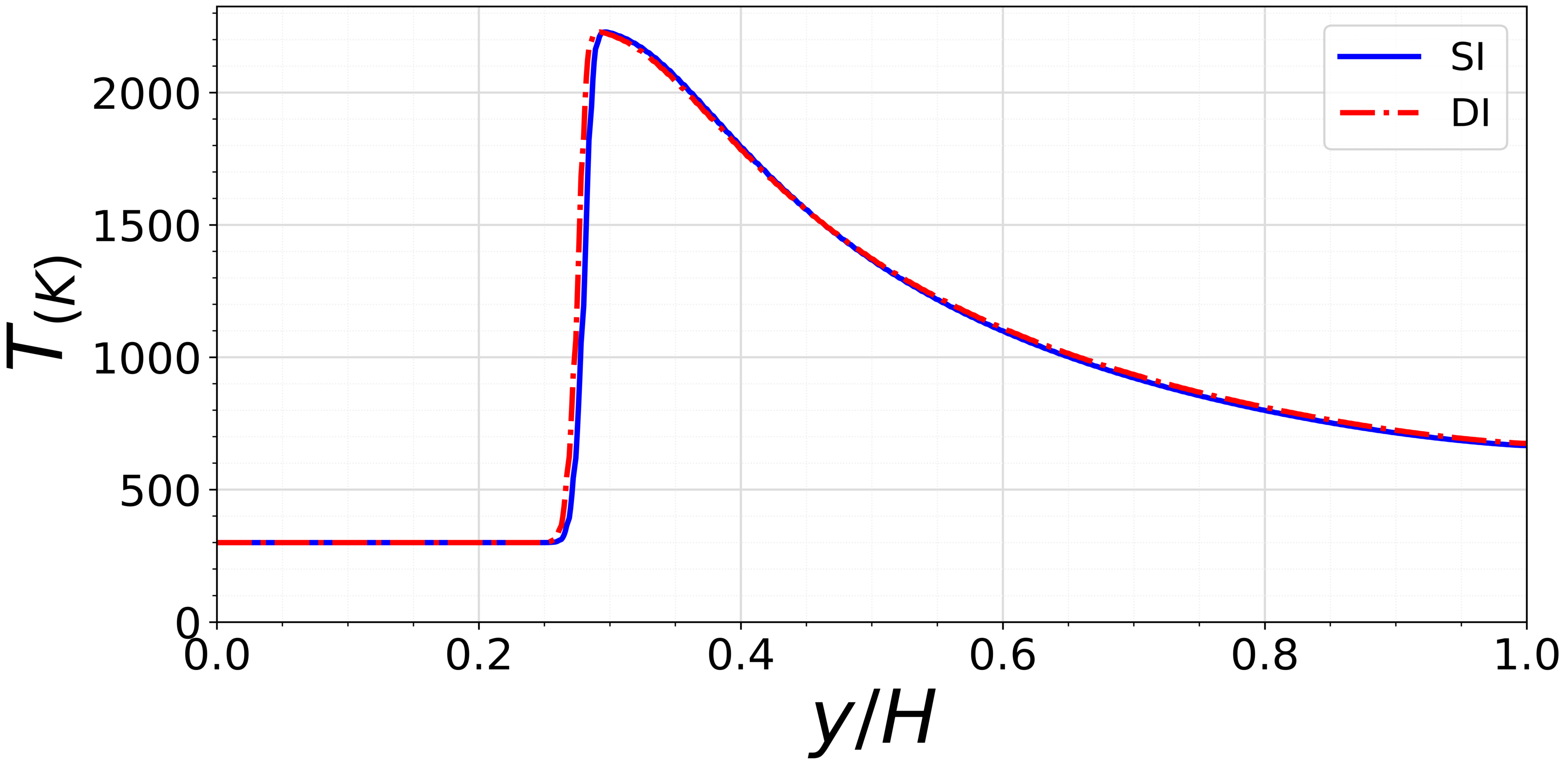}}
    \hideimage{\includegraphics[width=0.7\linewidth]{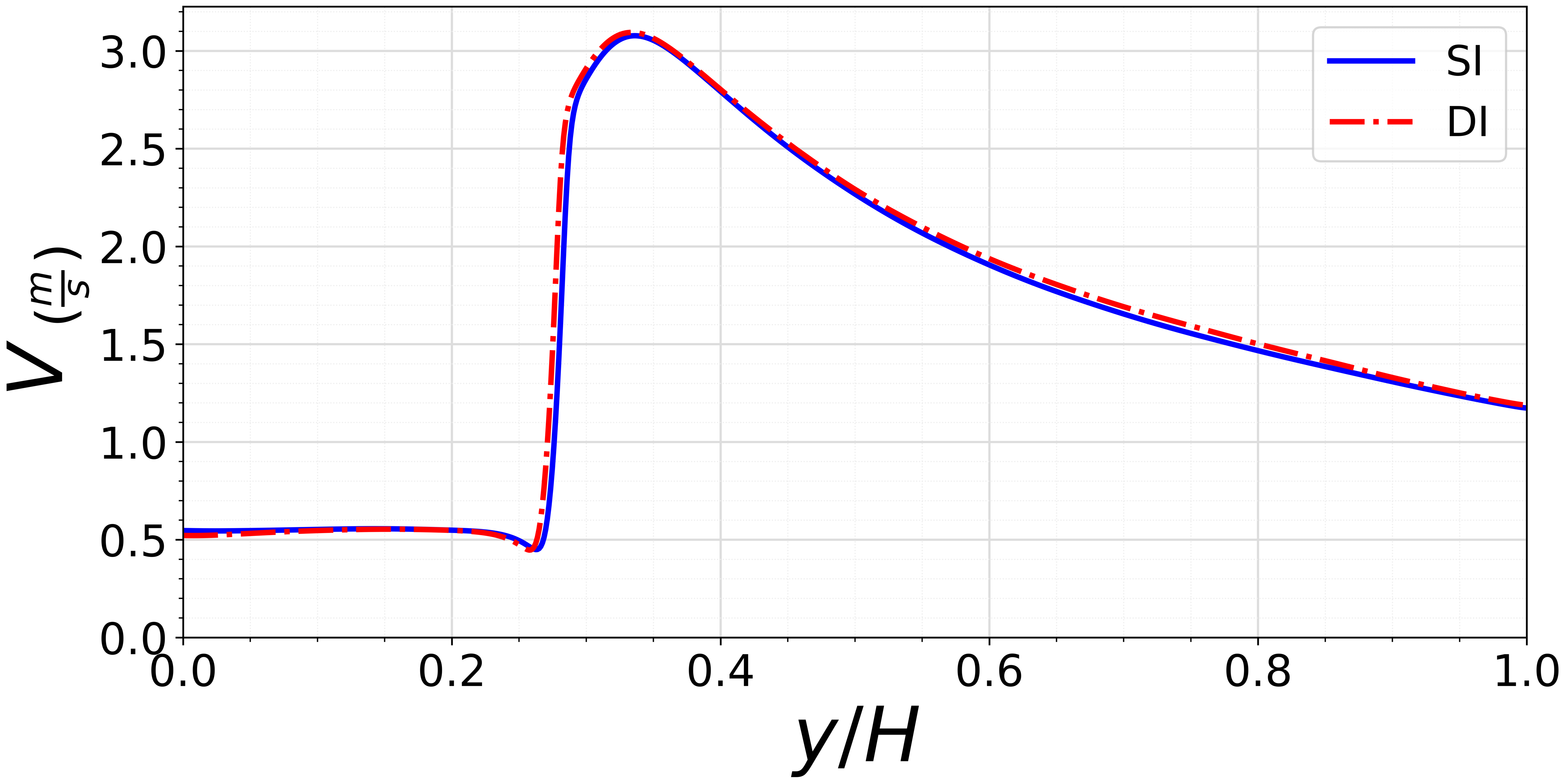}}
    \caption{A comparison of the results obtained within DI and SI-simulations for temperature and velocity magnitude --- averaged over the cross section --- versus the distance $y$ from the inlet. The values of the coupling parameters used in the diffuse interface formulation were  $\Austar=145$ and $\ATstar=70$. The result delivered by the DI-approach agrees well with SI-solution in the entire simulation domain.}
	\label{fig:SI-vs-DI-2-D-flame}
\end{figure}

To provide a further benchmark, we make use of the fact that results obtained within the proposed diffuse interface approach must be identical to sharp-interface solutions. Along this line, we have performed simulations of a 2D flame in a setup similar to the one shown in the upper row of Fig.~\ref{fig:2D-flame-isowalls-all} both within the present DI-approach and using the SI-version of our numerical tool (for details of the SI-method, see page~\pageref{page:SI-simulation-method}). Results obtained from these simulations are displayed in Figs.~\ref{fig:T-V-Flame-2D-DI}a and~\ref{fig:T-V-Flame-2D-DI}b for temperature and velocity fields across the entire channel. Further, Fig.~\ref{fig:SI-vs-DI-2-D-flame} provides a quantitative comparison between these diffuse and sharp interface simulations by showing the temperature and velocity magnitude over the cross section versus the longitudinal coordinate, $y$ (distance from the inlet). These comparisons show a good agreement between the DI- and SI-simulations and hence underline the reliability of the proposed diffuse interface methodology.

\subsubsection{Test against Experiment: Flame in a packed bed}
\label{sec:flame-in-packed-bed}
\begin{figure}[t]
	\centering
(a)    \hideimage{\includegraphics[height=4cm]{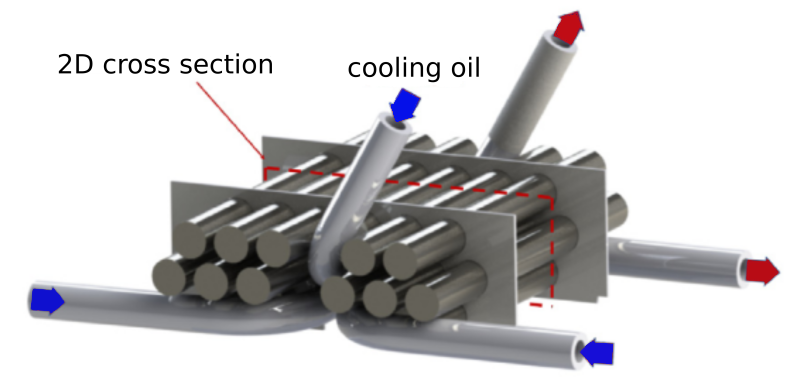}}
(b)    \hideimage{\includegraphics[height=4cm]{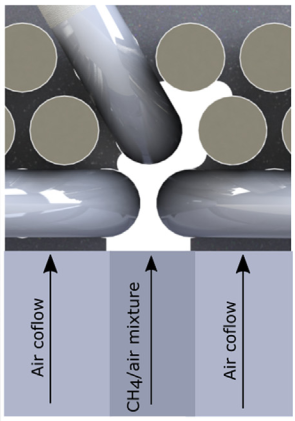}}
    \caption {Left: A schematic drawing of the setup used in experiments to study a stationary flame. This configuration was first used in~\cite{Khodsiani2023}. Here, cylinders are cooled to absorb the heat generated by chemical reactions. Since the mid-plane, which cuts the cylinders into two equal halves (red square) is relatively far from the borders, it is reasonable to assume that physical properties on this plane such as flow and temperature fields as well as chemical reaction processes are essentially independent of the axial position. This justifies the use of 2-D simulations of the corresponding cross sectional area. The right panel shows a zoom into the flow chamber and the inlet zone.}
	\label{fig:exp-setup}
\end{figure}

We proceed further with the validation of the proposed LB-FD-PF methodology and compare the simulated data with experiments on a stationary flame in a simple packed bed. The experimental setup consists of staggered arrays of cylinders and offers the possibility of direct visual monitoring of the position and shape of the flame (Fig.~\ref{fig:exp-setup}). The entire configuration is confined by parallel walls on the left and right. Each cylinder has a diameter of 10 mm, with a center-to-center distance of 12.3 mm. 

A methane-air mixture enters the packed bed from the bottom at a velocity of $\uinlet=0.3$m/s. The gas temperature at the inlet is $\Tinlet=300$K. The fuel mixture is ignited inside the chamber and the reaction products are transported away by the flow, exiting the channel from the top domain. The surface temperature of the three cylinders, which are arranged around the combustion zone,  is kept constant at $\Tcylinder=373$K. This is achieved by injecting cold silicon oil into the (hollow) cylinders. For further details on the experimental setup, please refer to ~\cite{Khodsiani2023}.

\begin{figure}[t]
	\centering
	\hideimage{\includegraphics[width=0.9\linewidth]{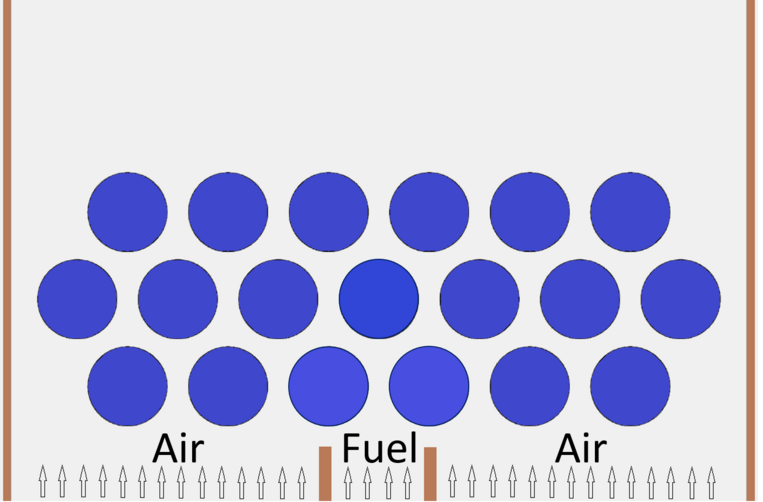}}
    \caption{A view of two-dimensional setup used in our simulation study of the packed bed of cylinders from Fig.~\ref{fig:exp-setup}. The main part of the inlet (bottom line in this plot) serves as entrance for air flow. The fuel (a mixture of methane and air) enters the domain only from a narrow region between the two filled squares (the "nozle"). Air and reaction products leave the simulation domain at the upper boundary.}
	\label{fig:sim-setup}
\end{figure}

It is important to note that, if one considers a cross sectional plane which cuts the cylinders into two parts of roughly the same length (see the dashed square in Fig.~\ref{fig:exp-setup}a), the flow and temperature fields as well as chemical reaction rates become approximately independent of the axial coordinate of the plane. This allows for a considerable reduction of the computation time by mapping the three dimensional problem onto a two dimensional one made up of an arrangement of circular discs (Fig.~\ref{fig:sim-setup}). The grid size and time step in our flame simulations are $\Delta x=0.025$ mm and $\Delta t=0.25\;\mu$s, respectively.

\begin{figure}[t]
	\centering
	\hideimage{\includegraphics[width=0.9\linewidth]{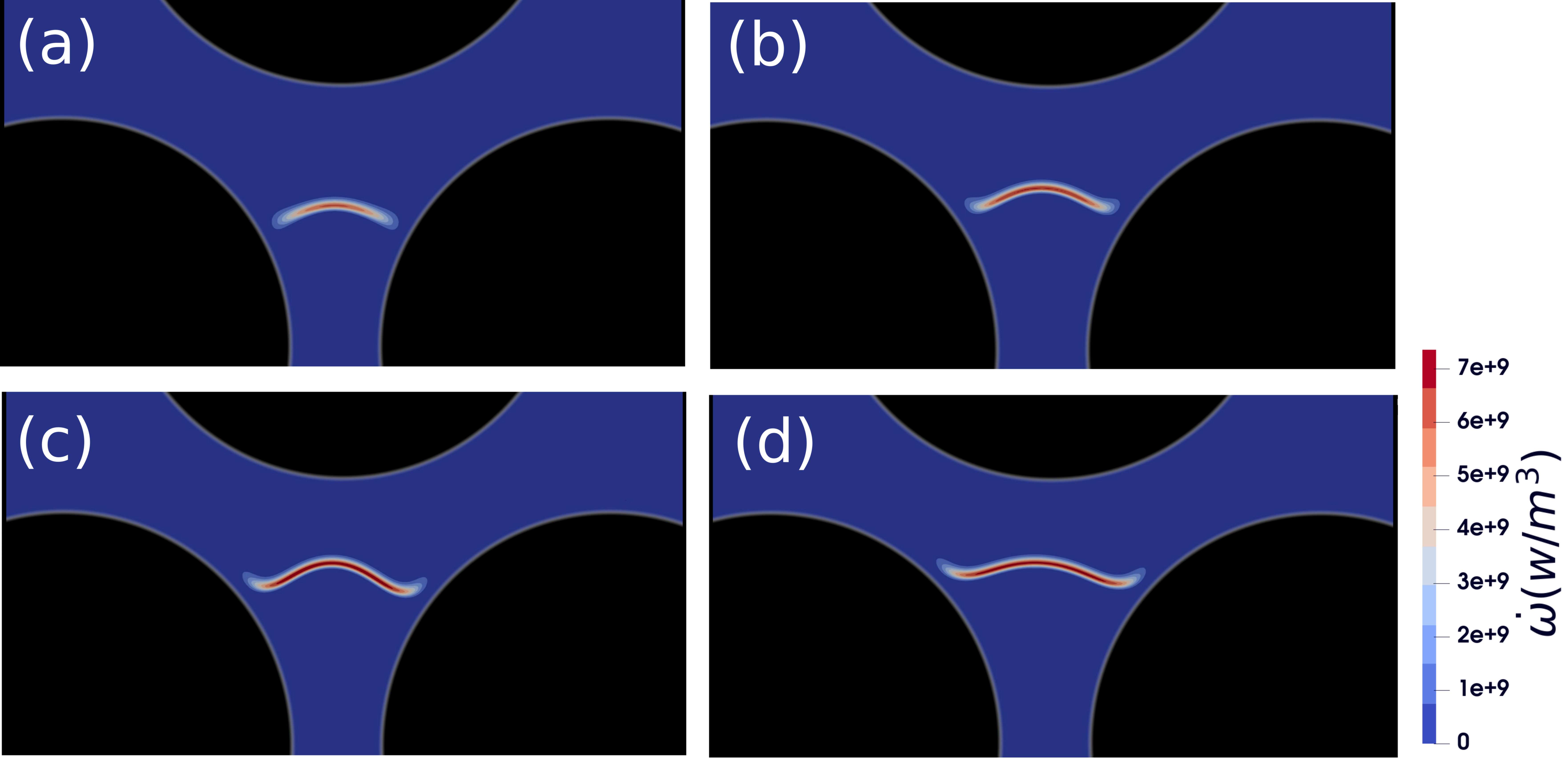}}
    \caption{Simulation results on the flame dynamics in a regular array of cylinders. The inlet fuel velocity is chosen such that the position and shape of the flame become stationary after a transient stage.}
	\label{fig:stablizing-flame}
\end{figure}

The process of flame formation and maintaining its stability requires particular attention in numerical simulations. In the simulation of gas flow in the packed bed, the heat release rate indicates the flame's location. At the beginning of this simulation, the zone between three specified cylinders is filled by burnt gas products and the upstream region is filled by fresh fuel mixture entering from the inlet. As depicted in Fig.~\ref{fig:stablizing-flame}, initially, a relatively weak flame emerges at the interface between fresh fuel and burnt products. This initial flame is transported by convection. However, over time, the interplay of convective, diffusive, and chemical reaction mechanisms, altering species concentrations, reaches a stationary state, allowing the flame to stabilize. It is worth mentioning that at the first time steps of the simulation, the temperature of the cylinders exposed to the flame is maintained above 373 K to create a warm environment necessary for the formation of the flame. Without this, the flame is not able to initiate.

As a first important quantitative validation against experiments, we have examined the condition for a stationary flame and find that simulations correctly reproduce a time-independent flame for the equivalence ratio used in experiments ($\phi=1$) if the fuel velocity at the inlet does not deviate from the experimental value ($\uinlet=0.3$m/s) by more than a few percent.

\begin{figure}[t]
	\centering
	\hideimage{\includegraphics[width=0.95\linewidth]{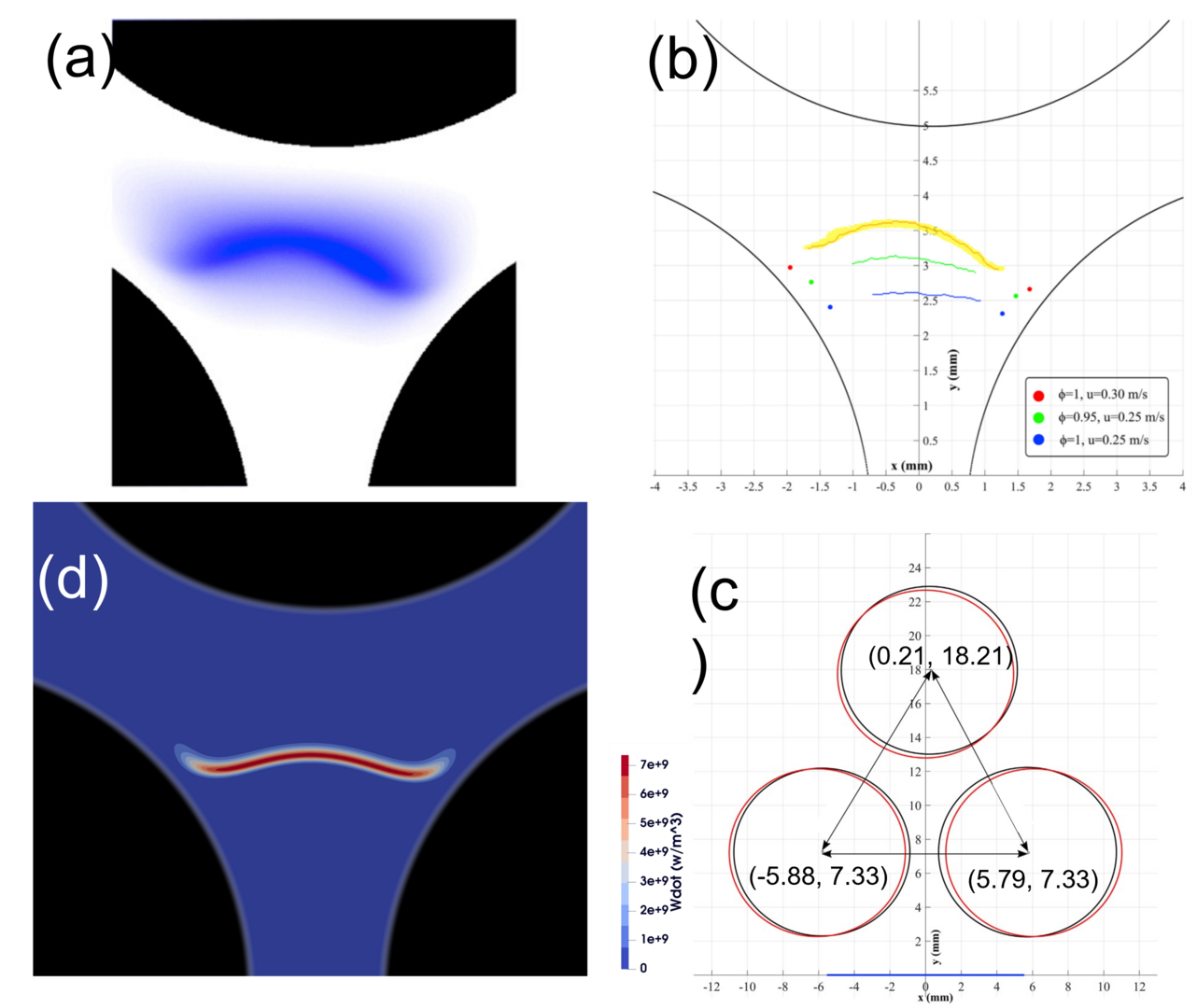}}
    \caption {Comparison between numerical simulations and experiments for a flame in a packed bed of cylinders. An image of the experimental flame is shown in the panel (a). Panel (b) shows the experimentally observed flames for a number of other inlet velocities and equivalence ratios as indicated in the legends. Panel (c) shows a cross sectional view of the actual cylinder arrangement in experiments (black) and a perfectly symmetric setup (red). The positions of cylinder centers are also given in this image. A segment of the inlet boundary, through which the methane-air mixture is injected into the channel (the \enquote{nozzle}) is also shown as a blue horizontal line. The simulated data is shown in panel (d), displaying the heat release rate, which serves here to highlight the position of the flame. Both in experiment and simulation, the inlet velocity is $\uinlet=0.3$m/s, the equivalence ratio is chosen to be $\phi=1$ and the temperature of the three cylinders, which surround the flame is kept constant at $\Tcylinder=373$K.  Panels (a), (b) and (d) are reproduced with permission (and with slight adaptation) from Particuology {\bf 85}, 167 (2023), Copyright 2023 Elsevier~\cite{Khodsiani2023}}.
	\label{fig:flame-sim-exp-comp}
\end{figure}

Further, we have compared the position and shape of the simulated flame with the experimental observations (Fig.~\ref{fig:flame-sim-exp-comp}). First we note that the experimental images of the flame show a small but discernible tilt (Fig.~\ref{fig:flame-sim-exp-comp}a) \cite{Khodsiani2023}. This tilt is systematic since it is also present for two other choices of inlet flow velocity and equivalence ratio (Fig.~\ref{fig:flame-sim-exp-comp}b). As a key property, this image also shows the average flame position and its temporal fluctuations (highlighted as a thick color band). 

The orign of the observed tilt of the flame lies in slight deviations of the arrangement of cylinders from a perfectly symmetric setup (Fig.~\ref{fig:flame-sim-exp-comp}c). These deviations are presumably due to limitations or imperfections in the manufacturing process and are therefore difficult to avoid. To mimic the experimental conditions as accurately as possible, we have replicated the experimental setup in our simulations by shifting the centers of the particles (discs) from their ideally symmetric positions in accordance with the experimental design. As seen in Fig.~\ref{fig:flame-sim-exp-comp}d, simulations of this setup also show the expected tilt. 

The above shown agremments between simulations and experiment on the flame propagation speed and on the shape of the flame front  highlight the reliability of the proposed simulation approach.

\hide{\begin{figure}[t]
	\centering
	\hideimage{\includegraphics[width=0.9\linewidth]{Figures/Exp-deviation.png}}
	\caption{\blue{}}
	\label{fig:exp-deviation}
\end{figure}
}

\hide{
	Figure~\ref{fig:vel-Temp-packed-bed} depicts the temperature and velocity fields the packed bed. It is evident from the figure that the velocity and temperature after passing through flame region increased significantly. The occurance of the two stagnation points of the velocity on the surface of two lower cylinders placed above the nozzle indicates that a portion of the injected mixture, after leaving the nozzle to the bed, tends to flow sideways. This observation ensures that the mixure encountering the flame was not mixed with airflow entering the bed from other parts of the inlet boundary. Consequently, the stoichiometric ratio of the mixture at the flame from remains at 1.
\begin{figure}[t]
	\centering
	\hideimage{\includegraphics[width=0.9\linewidth]{Figures/Temp-vel-packed-bed.png}}
	\caption{Temperature field...}
	\label{fig:vel-Temp-packed-bed}
\end{figure}
\begin{figure}[t]
	\centering
	\hideimage{\includegraphics[width=0.9\linewidth]{Figures/Temp-vel-packed-bed-zoom.png}}
	\caption{A zoom into the region....}
	\label{fig:vel-Temp-packed-bed-zoom}
\end{figure}
}

\subsubsection{Flame-front dynamics inside an assembly of complex-shaped particles}
\label{sec:flame-in-comlex-packed-bed}

\begin{figure}[t]
	\centering
	\hideimage{\includegraphics[width=0.9\linewidth]{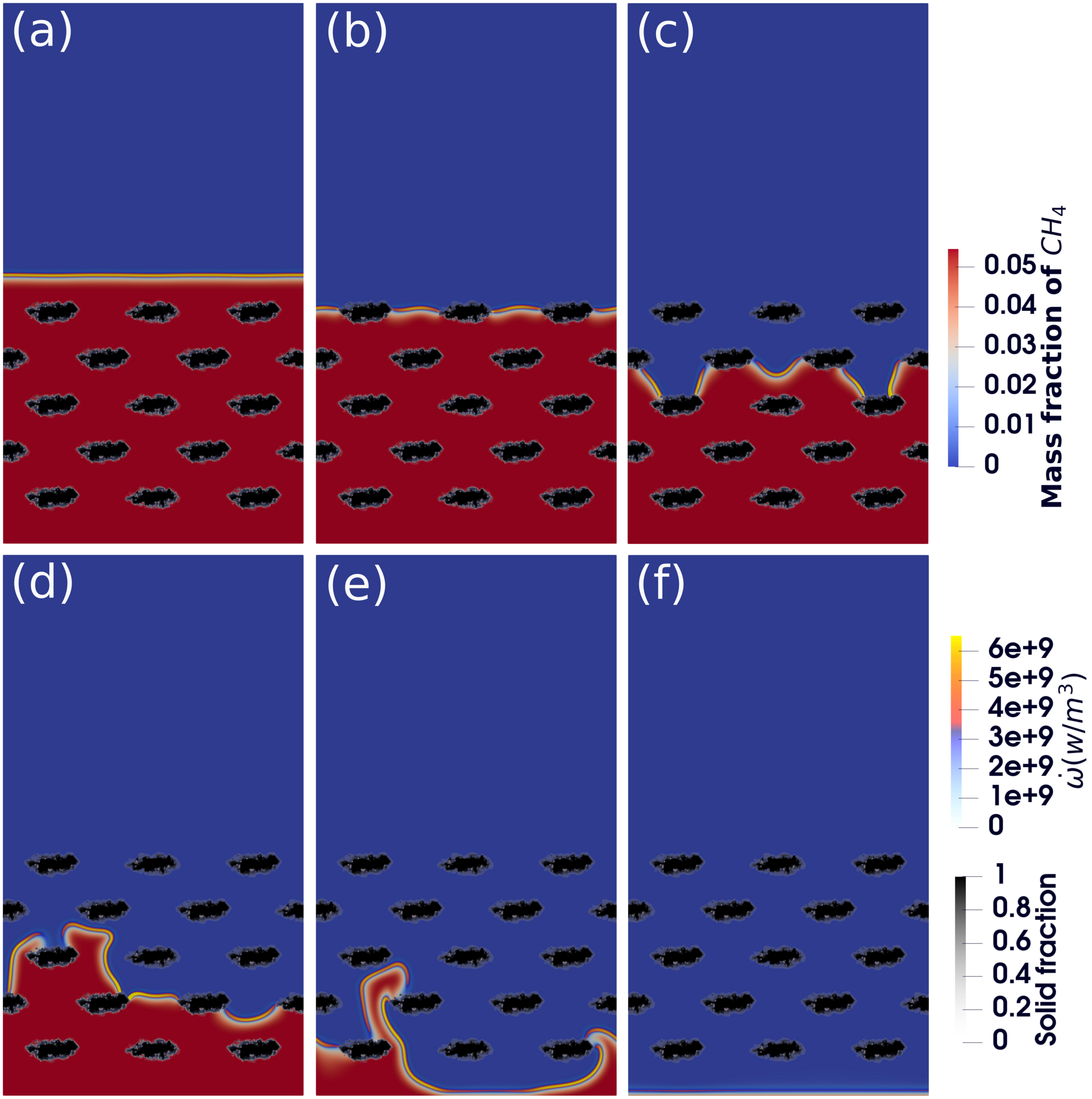}}
    \caption {Flame propagation through an assembly of particles with irregular shapes. These particles are generated by a grain growth model implemented in OpenPhase Academic (www.openphase.rub.de). Figures show mass fraction and heat release rate distributions of the methane and the flame within the bed. Peridic boundary condition was applied at side boundaries while solid bodies are insulated, keeping heat flux at the surfaces zero.}
	\label{fig:flame-complex-shape}
\end{figure}

As mentioned above, the phase field model is particularly suitable for studying complex shaped particles. This features is demonstrated in Fig.~\ref{fig:flame-complex-shape}, which shows flame propagation through an assembly of particles with non-regular shapes. The irregular shape used in these simulations is generated via phase field simulations of grain growth, combined with an empirical nucleation model. Subsequently, the coupled Phase Field-Lattice Boltzmann-Finite Difference simulations are performed to address reactive gas flow through this particle assembly. In this simulation setup, the lower half of the domain is filled with a feul/air mixture which is then ignited at the middle of the channel. The flow velocity at the inlet is chosen to be much lower than the estimated flame propagation speed. As a result, the flame front evolves toward the inlet. Periodic boundary condition is applied at lateral (left-right) borders of the simulation domain. The heat flux normal to the surface of the solid bodies is set to zero. In contrast to the isothermal boundary condition used in the above dicussed simulations, there is no separation zone between the flame and the solid body. Rather, the flame front is attached to the solid body and builds a normal angle with it. This behavior results from the fact that heat cannot flow through the solid body due to the imposed zero flux condition. Therefore, the flame orients itself such that the normal vector to its surface, along which energy can be transported most efficiently away from the flame, points parallel to the solid. As a result, the flame front itself takes a perpendicular alignment with respect to the solid surface.

\section{Conclusion}

In this study, we propose a systematic derivation of the momentum balance, heat and mass transport equations in the presence of a diffuse fluid-solid boundary. For this purpose, the corresponding partial differential equations, which are originally formulated for a sharp interface, are multiplied with a characteristic function, which indicates the presence of a given, say solid, phase and are then averaged over a small volume, whose size is dictated by the grid resolution. The resulting volume-averaged equations are then simplified by introducing appropriate interface quantities and neglecting terms of second order in fluctuations. The final coupled equations have a simple structure very much reminiscent of the original single-phase equations. The diffuse interface shows itself in coupling terms which act at the fluid-solid interface. 

The present hybrid lattice Boltzmann-finite difference-phase field (LB-FD-PF) scheme is thoroughly validated by conducting numerical simulations for a number of 1D and 2D test scenarios and comparing the results with sharp interface solution, available independent numerical data in the literature and carefully designed experiments. Last but not least, the proposed methodology is also applied to study flame propagation through an assembly of complex-shaped particles, thus highlighting the capability of the method in this case of great practical interest.

The proposed methodology opens the way for interesting new applications. On the one hand, the flexibility of the phase field method in generating complex structures is promising as it simplifies a realization of complex solid structures and their assemblies. The combination with the lattice Boltzmann-finite difference approach, on the other hand, allows to take adequate account of compression/expansion of the gas due to exchange/generation of heat. Therefore, the study of reactive gas flow through particle beds with complex shapes and arrangementss would be a particularly interesting application of the methodology proposed in this work. In the present study, the reaction is limited to the fluid phase. Including gas-solid reactions would be an interesting task for future work. In view of the versatility of the phase field method in dealing with phase transformation kinetics, an extension of this work towards chemical reactions at the fluid-solid contact appears very promising. Another very interesting application would be combustion of liquid fuel, which is often used in energy production. In such a situation, capillarity and contact line phenomena would come into play, which can be naturally treated in the present hybrid model thanks to the use of the phase-field approach.

\section*{Acknowledgement(s)}

This work has been funded by the Deutsche Forschungsgemeinschaft (DFG, German Research Foundation)--422037413-CRC/TRR 287 \enquote{BULK-REACTION}.

\section*{Disclosure statement}

No potential conflict of interest was reported by the author(s).

\section*{Funding}
This work was supported by the Deutsche Forschungsgemeinschaft (DFG, German Research Foundation) under Grant 422037413-CRC/TRR 287 \enquote{BULK-REACTION}.


\bibliographystyle{tfq}
\bibliography{refs}

\begin{thebibliography}{10}
\newcommand{\printfirst}[2]{#1}
\newcommand{\switchargs}[2]{#2#1}
\providecommand{\url}[1]{\normalfont{#1}}
\providecommand{\urlprefix}{Available at }

\bibitem{Anderson1984}
D.A. Anderson, \emph{Computational Fluid Mechanics and Heat Transfer},
  Hemisphere Publishing Corporation, New York, 1984.

\bibitem{Poinsot2005}
T. Poinsot and D. Veynante, \emph{Theoretical and numerical combustion}, RT
  Edwards, Inc., 2005.

\bibitem{Shaikh2018}
J. Shaikh, A. Sharma, and R. Bhardwaj, \emph{On comparison of the
  sharp-interface and diffuse-interface level set methods for 2d capillary
  or/and gravity induced flows}, Chem. Eng. Sci. 176 (2018), pp. 77--95.

\bibitem{Anderson1998}
G.B.M. D. M.~Anderson and A.A. Wheeler, \emph{Diffuse-interface methods in
  fluid mechanics}, Annu. Rev. Fluid Mech. 30 (1998), pp. 139--165.

\bibitem{Worner2012}
M. W{\"o}rner, \emph{Numerical modeling of multiphase flows in microfluidics
  and micro process engineering: a review of methods and applications},
  Microfluid. Nanofluid. 12 (2012), pp. 841--886.

\bibitem{Liu2022}
J. Liu, C. Huang, Z. Chai, and B. Shi, \emph{A diffuse-interface lattice
  boltzmann method for fluid--particle interaction problems}, Computers \&
  Fluids 233 (2022), p. 105240.

\bibitem{Liu2025}
X. Liu, C. Zhan, Y. Chen, Z. Chai, and B. Shi, \emph{A consistent and
  conservative diffuse-domain lattice boltzmann method for multiphase flows in
  complex geometries}, SIAM Journal on Scientific Computing 47 (2025), pp.
  B308--B332.

\bibitem{Varnik2007}
F. Varnik, D. Dorner, and D. Raabe, \emph{Roughness-induced flow instability: a
  lattice boltzmann study}, J. Fluid Mech. 573 (2007), p. 191–209.

\bibitem{Varnik2008}
F. Varnik, P. Truman, B. Wu, P. Uhlmann, D. Raabe, and M. Stamm, \emph{Wetting
  gradient induced separation of emulsions: A combined experimental and lattice
  boltzmann computer simulation study}, Phys. Fluids 20 (2008), p. 072104.

\bibitem{Ayodele2011}
S.G. Ayodele, F. Varnik, and D. Raabe, \emph{Lattice boltzmann study of pattern
  formation in reaction-diffusion systems}, Phys. Rev. E 83 (2011), p. 016702.

\bibitem{Ayodele2015}
S.G. Ayodele, D. Raabe, and F. Varnik, \emph{Shear-flow-controlled mode
  selection in a nonlinear autocatalytic medium}, Phys. Rev. E 91 (2015), p.
  022913.

\bibitem{Schiedung2017}
R. Schiedung, R.D. Kamachali, I. Steinbach, and F. Varnik,
  \emph{Multi-phase-field model for surface and phase-boundary diffusion},
  Phys. Rev. E 96 (2017), p. 012801.

\bibitem{Schiedung2018}
R. Schiedung, I. Steinbach, and F. Varnik, \emph{Multi-phase-field method for
  surface tension induced elasticity}, Phys. Rev. B 97 (2018), p. 035410.

\bibitem{Schiedung2020}
R. Schiedung, M. Tegeler, D. Medvedev, and F. Varnik, \emph{Simulation of
  capillary-driven kinetics with multi-phase-field and lattice boltzmann
  method}, Modell. Simul. Mater. Sci. Eng. 28 (2020), p. 065008.

\bibitem{Vakili2020}
S. Vakili, I. Steinbach, and F. Varnik, \emph{Multi-phase-field simulation of
  microstructure evolution in metallic foams}, Sci. Rep. 10 (2020), pp. 1--12.

\bibitem{Namdar2023}
R. Namdar, M. Khodsiani, H. Safari, T. Neeraj, S.A. Hosseini, F. Beyrau, B.
  Fond, D. Thévenin, and F. Varnik, \emph{Numerical study of convective heat
  transfer in static arrangements of particles with arbitrary shapes: A
  monolithic hybrid lattice boltzmann-finite difference-phase field solver},
  Particuology  (2023).

\bibitem{Drew1983}
D.A. Drew, \emph{Mathematical modeling of two-phase flow}, Annu. Rev. Fluid
  Mech. 15 (1983), pp. 261--291.

\bibitem{Beckermann1999}
C. Beckermann, H.J. Diepers, I. Steinbach, A. Karma, and X. Tong,
  \emph{Modeling melt convection in phase-field simulations of solidification},
  J. Comput. Phys. 154 (1999), pp. 468--496.

\bibitem{Subhedar2015}
A. Subhedar, I. Steinbach, and F. Varnik, \emph{Modeling the flow in diffuse
  interface methods of solidification}, Phys. Rev. E 92 (2015), p. 023303.

\bibitem{Subhedar2020}
A. Subhedar, P.K. Galenko, and F. Varnik, \emph{Thin interface limit of the
  double-sided phase-field model with convection}, Philosophical Transactions
  of the Royal Society A 378 (2020), p. 20190540.

\bibitem{Hosseini2019}
S.A. Hosseini, H. Safari, N. Darabiha, D. Th{\'e}venin, and M. Krafczyk,
  \emph{Hybrid lattice boltzmann-finite difference model for low mach number
  combustion simulation}, Combust. Flame 209 (2019), pp. 394--404.

\bibitem{Hosseini2020PhDTheis}
S.A. Hosseini, \emph{Development of a lattice boltzmann-based numerical method
  for the simulation of reacting flows}, Ph.D. diss., Universit{\'e}
  Paris-Saclay; Otto-von-Guericke-Universit{\"a}t Magdeburg,  2020.

\bibitem{Hosseini2020}
S.A. Hosseini, A. Abdelsamie, N. Darabiha, and D. Th{\'e}venin, \emph{Low-mach
  hybrid lattice boltzmann-finite difference solver for combustion in complex
  flows}, Physics of Fluids 32 (2020), p. 077105.

\bibitem{poinsot1992}
\emph{Boundary conditions for direct simulations of compressible viscous
  flows}, J. Comput. Phys. 101 (1992), pp. 104--129.

\bibitem{incropera1996}
F.P. Incropera, D.P. DeWitt, T.L. Bergman, A.S. Lavine, \emph{et~al.},
  \emph{Fundamentals of heat and mass transfer}, Vol.~6, Wiley New York, 1996.

\bibitem{Aydin2007}
O. Aydın and M. Avcı, \emph{Analysis of laminar heat transfer in
  micro-poiseuille flow}, Int. J. Therm. Sci. 46 (2007), pp. 30--37.

\bibitem{shah2003}
R.K. Shah and D.P. Sekulic, \emph{Fundamentals of heat exchanger design}, John
  Wiley \& Sons, 2003.

\bibitem{white1990}
F.M. White, \emph{Fluid mechanics}, New York, 1990.

\bibitem{kee2005}
R.J. Kee, M.E. Coltrin, and P. Glarborg, \emph{Chemically reacting flow: theory
  and practice}, John Wiley \& Sons, 2005.

\bibitem{franzelli2012}
B. Franzelli, E. Riber, L.Y. Gicquel, and T. Poinsot, \emph{Large eddy
  simulation of combustion instabilities in a lean partially premixed swirled
  flame}, Combust. Flame 159 (2012), pp. 621--637.

\bibitem{kim2006}
N.I. Kim and K. Maruta, \emph{A numerical study on propagation of premixed
  flames in small tubes}, Combust. Flame 146 (2006), pp. 283--301.

\bibitem{Hosseini2023}
S.A. Hosseini and D. Th{\'{e}}venin, \emph{Toward pore-scale simulation of
  combustion in porous media using a low-mach hybrid lattice
  boltzmann/finite-difference solver}, Phys. Fluids 35 (2023).

\bibitem{schlaffer2013}
M.B. Schlaffer, \emph{Non-reflecting boundary conditions for the lattice
  boltzmann method}, Ph.D. diss., Technische Universit{\"a}t M{\"u}nchen,
  2013.

\bibitem{Khodsiani2023}
M. Khodsiani, R. Namdar, F. Varnik, F. Beyrau, and B. Fond, \emph{Spatially
  resolved investigation of flame particle interaction in a two dimensional
  model packed bed}, Particuology  (2023).

\end{thebibliography}

\end{document}